\documentclass[a4paper,12pt]{article}
\pdfoutput=1
\usepackage{amsmath,amsfonts,amssymb,epsfig,comment,xspace,listings,hyperref,cite}
\usepackage{graphicx,caption,subcaption}

\numberwithin{equation}{section}
\usepackage{slashed}

\usepackage{youngtab}
\usepackage{listings}
\usepackage{xcolor}
\usepackage{booktabs}

\makeatletter

\def\pd[#1]{\frac{\partial}{\partial #1}}
\def\pdd[#1,#2]{\frac{\partial #1}{\partial #2}}

\def\nn{\nonumber}

\def\beq{\begin{equation}}
\def\eeq{\end{equation}}

\def\lagr{\mathcal{L}}

\def\twomat[#1,#2][#3,#4]{\left( \begin{array}{cc} #1 & #2 \\ #3 & #4 \end{array} \right)}
\def\twoa[#1,#2][#3,#4]{\left( \begin{array}{cc} #1 & #2 \\ #3 & #4 \end{array} \right)}

\def\thv[#1,#2,#3]{\left( \begin{array}{c} #1 \\ #2 \\ #3 \end{array} \right)}
\def\twv[#1,#2]{\left( \begin{array}{c} #1 \\ #2 \end{array} \right)}

\def\TeV{\ensuremath{\mathrm{TeV}}\xspace}

\def\SARAH{{\tt SARAH}\xspace}
\def\SPheno{{\tt SPheno}\xspace}
\def\ov{\overline}

\definecolor{maroon}{cmyk}{0, 0.87, 0.68, 0.32}
\definecolor{halfgray}{gray}{0.55}
\definecolor{slha_frame}{RGB}{207, 207, 207}
\definecolor{slha_bg}{RGB}{247, 247, 247}
\definecolor{slha_red}{RGB}{186, 33, 33}
\definecolor{slha_green}{RGB}{0, 128, 0}
\definecolor{slha_cyan}{RGB}{64, 128, 128}
\definecolor{slha_purple}{RGB}{170, 34, 255}

\definecolor{mathematica_frame}{RGB}{207, 207, 207}
\definecolor{mathematica_bg}{RGB}{247, 247, 247}
\definecolor{mathematica_red}{RGB}{186, 33, 33}
\definecolor{mathematica_green}{RGB}{0, 128, 0}
\definecolor{mathematica_cyan}{RGB}{64, 128, 128}
\definecolor{mathematica_purple}{RGB}{170, 34, 255}

\lstnewenvironment{MIN}[1][]{%
  \renewcommand{\thelstnumber}{In[\arabic{lstnumber}]}
  \lstset{language=MathIn,numbers=left,basicstyle=\ttfamily,#1}%
}{%
}

\lstnewenvironment{MOUT}[1][]{%
  \renewcommand{\thelstnumber}{Out[\arabic{lstnumber}]}
  \lstset{language=MathOut,numbers=left,basicstyle=\ttfamily,#1}%
}{%
}

\usepackage{listings}
\lstdefinelanguage{SLHA}{
    morekeywords={block,Block,BLOCK,decay,Decay,DECAY},%
    sensitive=true,%
    morecomment=[l]\#,%
    morestring=[b]',%
    morestring=[b]",%
    morestring=[s]{'''}{'''},
    morestring=[s]{"""}{"""},
    morestring=[s]{r'}{'},
    morestring=[s]{r"}{"},%
    morestring=[s]{r'''}{'''},%
    morestring=[s]{r"""}{"""},%
    morestring=[s]{u'}{'},
    morestring=[s]{u"}{"},%
    morestring=[s]{u'''}{'''},%
    morestring=[s]{u"""}{"""},%
    identifierstyle=\color{black}\ttfamily,
    commentstyle=\color{slha_cyan}\ttfamily,
    stringstyle=\color{slha_red}\ttfamily,
    keepspaces=true,
    showspaces=false,
    showstringspaces=false,
    rulecolor=\color{slha_frame},
    frame=single,
    frameround={t}{t}{t}{t},
    framexleftmargin=6mm,
    numbers=left,
    numberstyle=\tiny\color{halfgray},
    backgroundcolor=\color{slha_bg},
    basicstyle=\footnotesize,
    keywordstyle=\color{slha_green}\ttfamily,
    aboveskip=1.2em,
    belowskip=1.2em,
}

\lstdefinelanguage{MathIn}{
    morekeywords={Simplify,Eigenvalues},%
    emph={Start,InitUnitarity,GetScatteringDiagrams,BuildScatteringMatrix,MakeSPheno},%
    emphstyle={\color{mathematica_purple}},
    sensitive=true,%
    morecomment=[l]\%,%
    morestring=[b]',%
    morestring=[b]",%
    morestring=[s]{'''}{'''},
    morestring=[s]{"""}{"""},
    morestring=[s]{r'}{'},
    morestring=[s]{r"}{"},%
    morestring=[s]{r'''}{'''},%
    morestring=[s]{r"""}{"""},%
    morestring=[s]{u'}{'},
    morestring=[s]{u"}{"},%
    morestring=[s]{u'''}{'''},%
    morestring=[s]{u"""}{"""},%
    identifierstyle=\color{black}\ttfamily,
    commentstyle=\color{mathematica_cyan}\ttfamily,
    stringstyle=\color{mathematica_red}\ttfamily,
    keepspaces=true,
    showspaces=false,
    showstringspaces=false,
    rulecolor=\color{mathematica_frame},
    frame=single,
    frameround={t}{t}{t}{t},
    framexleftmargin=10mm,
    numbers=left,
    numberstyle=\tiny\color{halfgray},
    backgroundcolor=\color{mathematica_bg},
    basicstyle=\footnotesize,
    keywordstyle=\color{mathematica_green}\ttfamily,
    aboveskip=1.2em,
    belowskip=1.2em,
}

\lstdefinelanguage{MathOut}{
    morekeywords={Simplify,Eigenvalues},%
    sensitive=true,%
    morecomment=[l]\%,%
    morestring=[b]',%
    morestring=[b]",%
    morestring=[s]{'''}{'''},
    morestring=[s]{"""}{"""},
    morestring=[s]{r'}{'},
    morestring=[s]{r"}{"},%
    morestring=[s]{r'''}{'''},%
    morestring=[s]{r"""}{"""},%
    morestring=[s]{u'}{'},
    morestring=[s]{u"}{"},%
    morestring=[s]{u'''}{'''},%
    morestring=[s]{u"""}{"""},%
    identifierstyle=\color{black}\ttfamily,
    commentstyle=\color{mathematica_cyan}\ttfamily,
    stringstyle=\color{mathematica_red}\ttfamily,
    keepspaces=true,
    showspaces=false,
    showstringspaces=false,
    rulecolor=\color{mathematica_frame},
    frame=single,
    frameround={t}{t}{t}{t},
    framexleftmargin=10mm,
    numbers=left,
    numberstyle=\tiny\color{halfgray},
    backgroundcolor=\color{mathematica_bg},
    basicstyle=\footnotesize,
    keywordstyle=\color{mathematica_green}\ttfamily,
    aboveskip=1.2em,
    belowskip=1.2em,
}

\let\origthelstnumber\thelstnumber
\makeatletter
\newcommand*\Suppressnumber{%
  \lst@AddToHook{OnNewLine}{%
    \let\thelstnumber\relax%
     \advance\c@lstnumber-\@ne\relax%
    }%
}

\newcommand*\Reactivatenumber{%
  \lst@AddToHook{OnNewLine}{%
   \let\thelstnumber\origthelstnumber%
   \advance\c@lstnumber\@ne\relax}%
}

\numberwithin{equation}{section}

\title{How heavy can dark matter be? Constraining colourful unitarity with SARAH}

\date{}

\begin{document}

\begin{flushright}
\end{flushright}
\begin{center}

\vspace{1cm}
{{\bf \LARGE How heavy can dark matter be?} \\ {\bf \Large Constraining colourful unitarity with SARAH}}

\vspace{1cm}

\large{Mark D. Goodsell\footnote{goodsell@lpthe.jussieu.fr} and
Rhea Moutafis\footnote{moutafis@lpthe.jussieu.fr}
 \\[5mm]}

{ \sl Laboratoire de Physique Th\'eorique et Hautes Energies (LPTHE),\\ UMR 7589,
Sorbonne Universit\'e et CNRS, 4 place Jussieu, 75252 Paris Cedex 05, France.}

\end{center}
\vspace{0.7cm}

\abstract{We describe the automation of the calculation of perturbative unitarity constraints including scalars that have colour charges, and its release in \SARAH\ {\tt 4.14.4}. We apply this, along with vacuum stability constraints, to a simple dark matter model with colourful mediators and interesting decays, and show how it leads to a bound on a thermal relic dark matter mass well below the classic Griest-Kamionkowski limit.}

\newpage
\setcounter{footnote}{0}

\section{Introduction}

Unitarity of scattering amplitudes has long been used to constrain the masses and couplings of thermal relic dark matter (DM) particles \cite{Griest:1989wd,Hedri:2014mua,vonHarling:2014kha,Cahill-Rowley:2015aea,Kahlhoefer:2015bea,Baldes:2017gzw,ElHedri:2017nny,ElHedri:2018atj,Harz:2018csl,Hektor:2019ote,Kannike:2019mzk,Alanne:2020jwx,Fuks:2020tam,Espinoza:2020qyf,Espinoza:2020kut}. More generally, it is applied to constrain new physics Beyond the Standard Model such as $Z^\prime$ couplings \cite{Hosch:1996wu,Shu:2007wg,Babu:2011sd,Kahlhoefer:2015bea,Fuks:2020tam}, and most often (and relevant for this work) scalar couplings \cite{Lee:1977eg,Casalbuoni:1986hy,Casalbuoni:1987eg,Maalampi:1991fb,Kanemura:1993hm,Cynolter:2004cq,Schuessler:2007av,SchuesslerThesis,Kang:2013zba,Betre:2014fva,Costa:2014qga,Ginzburg:2003fe,Akeroyd:2000wc,Horejsi:2005da,Khan:2016sxm,Aoki:2007ah,Hartling:2014zca,DiLuzio:2016sur,DiLuzio:2017tfn,Goodsell:2018tti,Goodsell:2018fex,Krauss:2018orw,Espinoza:2018itz,Arhrib:2018sbz,Abbas:2018pfp,Cheng:2018mkc,Cheng:2018mkc,Chen:2018uim,Hektor:2019ote,Mondal:2019ldc,Kannike:2019mzk,Dubinin:2019dtb,Capdevilla:2020qel,Domenech:2020yjf,Alanne:2020jwx,Espinoza:2020qyf,Espinoza:2020kut} (including some one-loop calculations \cite{Grinstein:2015rtl,Cacchio:2016qyh,Murphy:2017ojk,Cheng:2018mkc}).

Unitarity famously limits the maximum possible cross-section for dark-matter annihilation, and thus gives an upper-bound on the mass of DM particles. The classic bound of ref.~\cite{Griest:1989wd} is derived for scattering momentum on-shell and represents a true all-orders bound, whereas standard constraints evaluated at large scattering momentum provide a complementary probe of the theory. Since they are usually evaluated at tree-level these should instead be considered really as a measure of the breakdown of perturbativity of the theory. 

To illustrate the relationship between the two, consider $2\rightarrow 2$ scattering processes from states $a \equiv (i,j)$ to $b \equiv(k,l)$ with matrix elements $\mathcal{M}_{ba}$ and centre-of-mass momenta $p_a, p_b$. We decompose them into partial waves with
\begin{align}
a_J^{ba} \equiv& \frac{1}{32\pi} \sqrt{\frac{4 |\mathbf{p}_{a}|\mathbf{p}_{b}|}{ 2^{\delta_a} 2^{\delta_b} s}} \int d z P_J (z) \mathcal{M}_{ba} (z)
\end{align}
where $\delta_{a} ( \delta_b)$ is $1$ for identical $i=j ( k=l)$ and $0$ otherwise;  
and $z$ the cosine of the angle between the three-momenta $\mathbf{p}_{a} , \mathbf{p}_{b}$. Then using unitarity of the corresponding S-matrix $S \sim 1 + i \mathcal{M}$, we find 
\begin{align}
\frac{1}{2i} (a_J - a_J^\dagger)^{ba} \ge \sum_c \ov{a}_J^{cb} a_J^{ca} \quad \forall a,b,J.
\end{align}
Since the matrix $a_J^{ba}$ is normal, we can diagonalise both sides simultaneously and so the same equation holds for the eigenvalues $a_J^{i}$; so the typical ``perturbative'' unitarity constraints yield
\begin{align}
|\mathrm{Re} (a_J^{i})| \le \frac{1}{2}.
\end{align}

To derive the limits of ref.~\cite{Griest:1989wd} we can invert the decomposition of partial waves and insert into the expression for the scattering cross-section $\sigma^{ba} = \sum_J \sigma_J^{ba}$ for states $a \rightarrow b$ to obtain:
\begin{align}
\sigma_J^{ba} =& 4\pi\frac{2J + 1}{p_a^2} 2^{\delta_a} |a_J^{ba}|^2.
\end{align}
Then we have
\begin{align}
\mathrm{Im} (a_J^{aa}) \ge |a_J^{aa}|^2 + |a_J^{ba}|^2 \longrightarrow |a_J^{ba}|^2 \le \frac{1}{4}
\end{align}
and this leads to an ``absolute'' bound\footnote{There are possible exceptions, such as in the presence of poles.} of
\begin{align}
\sigma_J^{ba} \le \pi\frac{2J + 1}{p_a^2} 2^{\delta_a} .
\end{align}
In limiting the dark matter mass, the factor of $ 2^{\delta_a}$ is compensated for non-identical particles by having two different species.

These bounds should be contrasted with the typical ``perturbative'' ones; for example, consider a toy model dark matter candidate $S$ with a $\mathbb{Z}_2$ symmetry that annihilates to a charged scalar $X$ via a quartic interaction:
\begin{align}
\mathcal{L}_{\rm toy} \supset - \frac{1}{2} \lambda_{\rm toy} S^2 |X|^2.
\end{align}
If we consider high-energy scattering as $s\rightarrow \infty$ then we obtain $a_0^{ba} = -\frac{\lambda_{\rm toy}}{16\pi \sqrt{2}}$ and we find the bound $\lambda_{\rm toy} < 8\pi \sqrt{2}$ \emph{at tree level}. This leads to the bound
\begin{align}
\sigma_0 \le 8\pi \frac{|\mathbf{p}_b|}{|\mathbf{p}_a| s}.
\end{align}
Consider now non-relativistic annihilation of the singlet $S$ into relativistic $X$, so $|\mathbf{p}_a| \approx m_S v, |\mathbf{p}_b| \approx m_S, s \approx 4 m_S^2, $ then we have the perturbative bound
\begin{align}
\sigma_0  \le \frac{2\pi}{m_S^2 v}  
\end{align}
compared to the ``absolute'' bound of
\begin{align}
\sigma_0 \le  \frac{2\pi}{m_S^2 v^2}.
\end{align}
Clearly even for this trivial case, for $v \ll 1$ the perturbative bound is stronger and will lead to a lower limit on the DM mass, since we have taken the bound on $\lambda_{\rm toy} $ at $s\rightarrow \infty$ and applied it for small $s$. Crucially, though, this bound is really a measure of the \emph{perturbativity} of the theory, since we only derived it with tree-level information, so it is entirely possible that a theory would saturate the ``absolute'' bound in the non-perturbative regime. 

In our toy example, we included for simplicity only a quartic coupling and took $s\rightarrow \infty$. This is rather typical in the literature among calculations of unitarity constraints. These ignore the contributions from, in particular, scalar trilinear couplings -- which have enormous implications for dark matter phenomenology, since they are responsible for all $s/t/u$ channel interactions. 
However, a framework within the package \SARAH \cite{Staub:2008uz,Staub:2013tta} for \emph{automatically} calculating the constraints on scalar trilinears was introduced in ref.~\cite{Goodsell:2018tti}, which can automatically scan over scattering momentum to find the best limit on the couplings of the theory. This has since been applied in e.g. ref.~\cite{Goodsell:2018fex,Krauss:2018orw,Espinoza:2018itz,Hektor:2019ote,Mondal:2019ldc,Kannike:2019mzk,Alanne:2020jwx,Espinoza:2020qyf,Espinoza:2020kut}. As we saw above even in a trivial example, this will lead to generally stronger bounds on the dark matter mass than in ref. \cite{Griest:1989wd}. However, the calculation in ref.~\cite{Goodsell:2018tti} was until now limited to \emph{colour neutral} scalars. In this paper we shall describe the extension in \SARAH\ {\tt v4.14.4} to \emph{colourful} scalars, where all group theory factors are automatically calculated, and use this to place constraints on scalar trilinear couplings that are relevant for a simple dark matter model with colourful mediators.

Unitarity, however, is not the only constraint on trilinear couplings: they can also lead to alternative vacua, which in the case of charged fields mean charge- or colour-breaking minima of the potential. These are offset by having larger quartic couplings to stabilise the vacuum at the origin in field space. The typical approach to constraining a new model with such scalars, therefore, would be to use vacuum stability to constrain the size of cubic couplings, which in turn push the theory to large quartic couplings; large scattering-momentum unitarity to give an upper bound on the quartic couplings; and the dark matter annihilation cross-section is then limited by the values of both (since it can proceed via both quartic and $s/t/u$-channel interactions).

This reasoning is reinforced, as discussed for example in ref.~\cite{Goodsell:2018tti}, by the fact that for a single neutral scalar field with both cubic and quartic couplings, the full bounds from unitarity on the cubic coupling are generally \emph{less} constraining that those from vacuum stability plus the upper limit on the quartic from unitarity. On the other hand, this naive picture does not necessarily hold for models with colourful states, or more scalars, but up until now there was no simple way of deriving the unitarity constraints for such theories. To our knowledge, such bounds had only been applied in a model with a colour octet in ref.~\cite{Cao:2013wqa,He:2013tla,Cheng:2018mkc}\footnote{We thank Junjie Cao for bringing the first of these to our attention after the first version of this paper.} (in the large scattering momentum limit only); and in the (N)MSSM in ref.~\cite{Schuessler:2007av,SchuesslerThesis} (with a scan over scattering momentum as discussed here) and \cite{Staub:2018vux} (using an earlier version of the code described in this paper). In the latter reference, a comparison of unitarity and vacuum stability bounds was performed for the Higgs-squark sector where the conclusion was that the unitarity constraints on the trilinear and quartic couplings between scalars were irrelevant in the MSSM (where the quartic couplings are given only by gauge and Yukawa couplings) but were \emph{complementary} to the vacuum stability constraint in the NMSSM. However, in those models the colourful scalar sectors interact only with the Higgs scalars, which cannot provide a dark matter candidate. We also point to ref.~\cite{Baker:2020vkh}, which makes use of the routines described here to constrain models of radiative fermion mass generation. 

In this paper we shall investigate in detail the (genuine) complementarity of the requirements of (full) unitarity including finite momentum scattering, vacuum stability and relic density to place an upper bound on a scalar dark matter model with colourful mediators for the first time, which will allow us to put an upper bound on the dark matter mass well below the Griest-Kamionkowski limit. In section \ref{SEC:Model} we describe our model and how we have calculated vacuum stability bounds for it; in sec.~\ref{SEC:ColourfulUnit} we describe the automatisation of the group theory calculations as we have implemented in \SARAH v4.14.4; in sec.~\ref{SEC:Results} we describe the procedure that we used to investigate the parameter space of our model and show the results, giving an upper bound on the mass of the dark matter particle.

\section{A model of colourful mediators}
\label{SEC:Model}

To illustrate the new capabilities in \SARAH and test the idea of a maximum dark matter mass, we shall take a model with colourful scalar mediators, but where the dark matter candidate is the usual scalar singlet $S$ with a $\mathbb{Z}_2$ symmetry. The scalar mediator fields $Q_E$ and $Q_O$ both have quantum numbers $(3,1)_{-1/3}$ under $(SU(3), SU(2))_Y$; the difference between them is that $Q_E$ is even under the $\mathbb{Z}_2$, and $Q_O$ is odd. Then the most general lagrangian where the hidden sector respects CP symmetry is 
\begin{align}
  \lagr =& \lagr_{SM} - \frac{1}{2} m_S^2 S^2 - m_E^2 |Q_E|^2 - m_O^2 |Q_O|^2 - \lambda_S S^4 - \frac{1}{2} \lambda_{HS} S^2 |H|^2 - \lambda_{3} |H|^2 |Q_E|^2 - \lambda_{4} |H|^2 |Q_O|^2 \nn\\
         & - \frac{1}{2} \lambda_1 S^2 |Q_O|^2 - \frac{1}{2} \lambda_2 S^2 |Q_E|^2 -  \lambda_5 |Q_E|^4  -  \lambda_6 |Q_O|^4 -  \lambda_7 |Q_O|^2 |Q_E|^2 - \lambda_8 |Q_O Q_E^*|^2 \nn\\
  & - \bigg[ \kappa_1 S Q_E Q_O^* + Y_Q^{ij} Q_E q_i q_j + \frac{1}{4} \lambda_C (Q_E Q_O^*)^2 + h.c. \bigg]
\end{align}
Here $q_i$ are the $(3,2)_{1/6}$ Weyl fermions representing left-handed SM quarks. 
This model has several interesting features. The first, which is the main point of considering it, is the trilinear coupling $\kappa$: this entirely controls the $s/t/u$-channel processes for dark-matter annihiliation and is crucial for the unitarity and vacuum stability analysis. The next is the baryonic coupling $Y_Q^{ij}$: the mediators carry baryon number, which is respected by the model (perturbatively). It also means that the state $Q_E$ decays to pairs of quarks; we shall take it to predominantly couple to the third generation, i.e. decays to a $tb$ pair. Therefore it is somewhat hard to search for at the LHC, being constrained mainly by $t \ov{t} b \ov{b}$ searches for which no BSM reanalysis is yet possible, so we expect its mass to be only bounded to be larger than $1$ TeV (rather than $2$ TeV and above for other colourful scalars that decay to the first two generations of quarks). This choice also makes the model somewhat safe from direct detection constraints (provided that the Higgs portal coupling $\lambda_{HS}$ is small). In this work, we shall be considering in any case much larger masses, so collider and direct searches are not relevant. 

Another interesting feature is that the state $Q_O$ can only decay to the singlet plus $Q_E$, requiring it to be heavier than the singlet.
In addition, there are three operators containing two pairs of $Q_O, Q_E$, namely the $\lambda_7, \lambda_8$ and $\lambda_C$ terms. It is now possible within \SARAH to specify all of these and for them to be properly taken into account in the unitarity constraints; however, for our analysis we shall only consider $\lambda_7$ and take $\lambda_C, \lambda_8$ to be zero. This is mildly relevant for unitarity and vacuum stability constraints -- but not at all for the dark matter density.

Since we are considering heavy dark matter that has little interaction via the Higgs portal, the relevant part of the scalar potential for this model involves the fields $S, Q_E$ and $Q_O$. These can develop expectation values and a colour-breaking minimum if $\kappa$ is large enough; however, finding the minimum of the potential involves solving coupled cubic equations and is not analytically tractable except for the the point where the masses and couplings are equal. To find possible true minima we wrote a small {\tt Python} code which we briefly describe in appendix \ref{APP:VacStab}. This uses {\tt HOM4PS2} \cite{hom4ps2} to quickly find \emph{all} minima of the set of coupled minimisation conditions for our chosen field directions. We found this simpler than installing the no-longer-supported {\tt Vevacious} \cite{Camargo-Molina:2013sta}, especially since there is a potentially large separation of scales between our dark matter sector and the Higgs sector; also note that we are only interested in the \emph{tree-level} minima because we are explicitly searching for points which have large trilinear couplings where perturbativity may break down.

\section{Colourful unitarity bounds}
\label{SEC:ColourfulUnit}

Unitarity bounds on colourful scattering amplitudes for the MSSM were considered in \cite{SchuesslerThesis} where a derivation of the colour factors was given case by case for the different representations and amplitudes present. Here we shall give a description of the general procedure that we use, that applies to the scattering of \emph{any} states. 

Let us suppose that our initial (or final) states can be labelled $A_i, B_j$ and transform non-trivially under a non-Abelian group, let us say with dimensions $d_A, d_B$. This means that we multiply the number of rows that it takes up in the scattering matrix by $d_A \times d_B$. Clearly, however, we can break this into irreducible representations:
\begin{align}
d_A \times d_B = \sum_{C}^{n} d_C ,
\end{align}
where $n$ is the total number of irreducible representations. 
Obviously the scattering matrix will only be non-zero when the incoming and outgoing pairs are in the same irrep, so then we need to apply a unitary transformation on the $d_A d_B$ states to split them into $n$ blocks; these are given by (generalised) Clebsch-Gordan coefficients. These can be built from invariant tensors, that is a mapping of $A\otimes B \otimes C^* \rightarrow 1$; we can denote this as  $(t_C)^{ij }_{a}$ so that $A_i B_j \ov{C}^{a} (t_C)^{ij}_{a}$ is invariant under group tranformations. By considering infinitesimal transformations it is easy to see that contracting different invariant tensors together make another invariant tensor, and since the only invariant with just one representation and its conjugate is a Kronecker delta, then we must have\footnote{See also ref.~\cite{Cao:2013wqa} for explicit Clebsch-Gordan coefficients for a model with octets.}
\begin{align}
  (t_C)^{ij }_{a} (\ov{t}_C)_{ij }^{b}  \propto& \delta_{a}^{b}.
\end{align}
However, there could be more than one copy of any given representation in the decomposition above --  the most relevant example here being for a product of two octet representations, for which
\begin{align}
\mathbf{8} \times \mathbf{8} = & \mathbf{1} + \mathbf{27} + \mathbf{10} + \mathbf{\ov{10}} + 2 \times  \mathbf{8}
\end{align}
where the relevant bit is the appearance of two $\mathbf{8}$ reps; this is more familiarly understood as the existence of two invariants, $d^{abc}$ and $f^{abc}$, which contract the symmetric and antisymmetric combinations. Hence if we have two or more copies of a given representation, we can label them $C$ and $D$ and have 
\begin{align}
(t_C)^{ij }_{a} (\ov{t}_D)_{ij }^{b}  =& g^{CD} \delta_{a}^{b}, \qquad \bigg(g^{CD} = 0 \ \mathrm{if\ reps}\ C,D\ \mathrm{not\ identical} \bigg)
\end{align}
Now we are free to diagonalise the basis of invariants and normalise them appropriately. 

Since the scattering matrix is an isomorphism of the initial to final colour rep, by Schur's lemma it is proportional to the identity. Then each matrix will just be $d_C$ copies of this along the diagonal. So then we need to do a unitary transformation $R_{ij,i'}$ on the scattering matrix to split it into blocks.
For it to be unitary, we need
\begin{align}
R^{ij}_{a} \ov{R}_{ij}^{b} =& \delta^{b}_{a} \delta_{C D}, \qquad a \in C, b \in D\label{EQ:rconds1}\\  
\sum_C \sum_{a \in C } R^{ij}_{a} \ov{R}_{kl}^{a} =& \delta^{i}_k \delta^{j}_l
\label{EQ:rconds2}\end{align}
Note that the second line involves \emph{the sum over all representations present}. From the above, it is clear that we can construct these matrices from our diagonalised basis of invariants, and the first condition means that we must take $g^{CD} = \delta^{CD}$ and $R^{ij}_{a} = \oplus_C (t_C)^{ij }_{a}$.

Translating this to amplitudes, for $i,j \rightarrow k,l$ we have a scattering matrix $\mathcal{M}^{kl}_{\ \ ij}$ or equivalently $(a_{0})^{kl}_{\ \ ij}$ upon which we are free to make unitary tranformations of the states to get
\begin{align}
  (\ov{t}_C)_{kl }^{b} (a_0)_{\ \ ij}^{kl} (t_C)^{ij }_{a}  \equiv \delta_a^b a_0^{(C)},
\end{align}
since outgoing states are equivalent to conjugated incoming ones. 
So, once we have constructed the invariants, we contract them with our scattering matrices to obtain a block-diagonal form. We now have a choice to extract $a_0^{(C)}$: we can take the trace over the remaining indices $a, b$, pick one example, or construct $a_0 a_0^\dagger$ on colour space and take the square root of the diagonal entries. In \SARAH we take the simplest choice and put $a = b =1$ as constraints in the evaluation of the amplitudes as it is by far the least computationally expensive. However, it should be noted that, if some of the couplings/invariants are specified by the user in a different basis, then there could in principle be a rotation between the incoming and outgoing states which would then yield incorrect results here.

\subsection{Examples}
\label{SEC:ColourExamples}

The general technique that we use here is different from the approach in ref.~\cite{SchuesslerThesis}, and so it is instructive to give some simple examples. We did cross-check all of the colour factors produced by the \SARAH in the (N)MSSM with the results there. However, since the colour representations available in those models are not different from ours, we instead give examples directly in the model here and in appendix \ref{APP:Storage}.  

Consider first our dark matter annihilation channel $S, S \rightarrow (Q_E)_i, (\ov{Q}_E)^j$. We can decompose the final state into a singlet and an octet, but here we can only find the singlet representation. To find the projectors we can consider the $SU(N)$ identity
\begin{align}
 \delta_{ii' } \delta^{jj'} =&  \frac{1}{N}  \delta_i^j \delta^{j'}_{i'} + 2 (T^a)_i^j (T^a)_{i'}^{j'}
\end{align}
the projectors for the singlet and octet  are $\frac{1}{\sqrt{3}} \delta_j^i$ and $\sqrt{2} (T^a)_j^i$ in order for the above equation to become equation (\ref{EQ:rconds2}). In our model we only have $t/u$-channel annihilation via $Q_O$ exchange, so the diagram is proportional to $\kappa_1^2 \delta_i^j $ and so
\begin{align}
a_0^{(0)} (S S \rightarrow Q_E \ov{Q}_E) \propto \kappa_1^2 \frac{1}{\sqrt{3}} \times 3 = \sqrt{3} \kappa_1^2.  
\end{align}
Similarly the $t/u$-channel elastic interaction $Q_E \ov{Q}_E \rightarrow Q_E \ov{Q}_E \propto 3 \kappa_1^2.$ 

Consider now the interaction with coupling $\lambda_5$ with scattering of $Q_E, \ov{Q}_E$ pairs to each other. The vertex in this case is $- 2\lambda_5 (\delta_i^j \delta_k^l + \delta_i^l \delta_k^j)$. %
So for this diagram via the singlet and octet channels we have
\begin{align}
  a_0^{(0)} (Q_E \ov{Q}_E \rightarrow Q_E \ov{Q}_E) \underset{s\rightarrow \infty}{=} & - 2\lambda_5 \frac{2}{32\pi} \frac{1}{3} ( 9+ 3) = - \frac{\lambda_5}{2\pi} \\
  a_0^{(8)} (Q_E \ov{Q}_E \rightarrow Q_E \ov{Q}_E) \underset{s\rightarrow \infty}{=} & - 2\lambda_5 \frac{2}{32\pi} 2 \mathrm{tr}(T^1 T^1) =  -  \frac{\lambda_5}{8\pi} 
\end{align}
Hence in the $s\rightarrow \infty$ limit we have the strongest limit from the singlet representation, and a limit of $\lambda_5 \le \pi$; the same limit applies for $\lambda_6$. 

If we consider $Q_E, Q_E$ scattering then we can use the same vertex, but now we decompose the representations into $\mathbf{3} + \mathbf{6}.$ The projector for the antisymmetric combination can be taken to be $\frac{1}{\sqrt{2}} (\delta_{1i} \delta_{2k} -\delta_{2k} \delta_{1i} )$ for incoming states (and $(i\leftrightarrow j, k \leftrightarrow l)$ for outgoing) and for the symmetric one we can just take $\delta_{1i} \delta_{1k}$ or equivalently $\frac{1}{\sqrt{2}} (\delta_{1i} \delta_{2k} +\delta_{2i} \delta_{1k} ).$ These lead to 
\begin{align}
 a_0^{(3)} (Q_E Q_E \rightarrow Q_E Q_E) \underset{s\rightarrow \infty}{=} & 0  \\
  a_0^{(6)} (Q_E Q_E \rightarrow Q_E Q_E) \underset{s\rightarrow \infty}{=} &  -  \frac{\lambda_5}{4\pi}.
\end{align}
Hence again these give weaker bounds than the singlet representation.


\section{Limiting the dark matter mass}
\label{SEC:Results}

Now that we have assembled the relevant machinery, in this section we will finally search for an upper bound on the dark matter mass in our model. To do this we use the \SPheno \cite{Porod:2003um,Porod:2011nf} code generated by \SARAH for our model to calculate the spectrum, decays and unitarity constraints; we use the vacuum stability code described in appendix \ref{APP:VacStab} to determine whether the colour-preserving vacuum is stable; and we use {\tt micrOMEGAs\ 5.2.1} \cite{Belanger:2018ccd,Belanger:2020gnr} to calculate the dark matter relic density and direct detection cross-sections. Since we are interested in the \emph{allowed} parameter space of the model, we will simply require that the dark matter relic density not exceed the Planck value $\Omega h^2 = 0.120 (3)$ \cite{Aghanim:2018eyx}. All constraints on the parameter space are listed in table~\ref{tab:Constraints}. 

\begin{table}[h]
\centering
\begin{tabular}{ll} 
  \toprule
  Dark matter density & $\rho_\text{DM} \geq \Omega h^2 = 0.120(3)$ \\ 
  Vacuum stability & $S\equiv x, Q_E \equiv \frac{1}{\sqrt{2}}y, Q_O \equiv \frac{1}{\sqrt{2}}z, \quad x, y, z \in \mathbb{R}$ \\
  Mass hierarchy and cubic coupling & $\kappa, m_E \leq m_S \leq m_O, \quad \text{where}\ m_S \lesssim \mathcal{O}(300 \text{TeV})$ \\
  Quartic couplings & $\Lambda \equiv \lambda_5 = \lambda_6 = 4\lambda_S \leq 3.5$ \\
  Only decay of $Q_E$ to top-bottom pair & $Y_Q^{33} = 1$, all other terms are 0 \\ [0.5ex]
  \bottomrule
\end{tabular}
\caption{Constraints on the allowed parameter space.}
\label{tab:Constraints}
\end{table}

However, to find the maximal dark matter mass with these constraints in our model with three heavy scalars, a cubic coupling and several quartic couplings involves a search on a multidimensional parameter space. We are interested in the mass hierarchy $m_O > m_S > m_E$ and in exploring ranges of $m_S$ up to $\mathcal{O} (300)$ TeV. Moreover, the quartic couplings should naively be bounded by $\lambda_S \le \frac{2 \pi}{3}, \lambda_{5,6} \le \pi  $. However, as seen in \cite{Goodsell:2018tti}, cancellations between the contributions from quartics and cubic couplings, and the effect of a finite momentum cutoff could in principle allow somewhat larger values. Therefore, in a series of Markov Chain Monte Carlo (MCMC) scans we explored larger values with the final, finer scan of one million points having an upper limit of $\lambda_{5,6} \le 3.5$. We chose this rather generous upper limit (instead of, say, $\lambda_{5,6} \le 3.2$) to make sure that there are no unexpected phenomena in a theoretically excluded range. These are the most important quartic couplings since they control the overall stability of the potential. As described in appendix \ref{APP:VacStab}, for our vacuum stability determination, we consider field directions along $S \equiv x, Q_E \equiv \frac{1}{\sqrt{2}} y, Q_O \equiv \frac{1}{\sqrt{2}} z $ where $x,y,z $ are real. We see that  taking $\lambda_5 = \lambda_6 = 4\lambda_S$ renders the potential symmetric in $x,y,z$ at large field values (when other couplings vanish) so for simplicity we impose this condition in our search which leaves us with a scan over 
\begin{align}
\kappa, m_E \le m_S, m_O \ge m_S, \Lambda \equiv \lambda_5 = \lambda_6 = 4 \lambda_S,
\end{align}
and we simply take the other quartic couplings to zero except for $\lambda_7$ which we, quite arbitrarily, set to 0.1 although this has no impact on the search, except perhaps a very slight influence on vacuum stability. In other words, we are allowing self-couplings of the mediators and the singlet, respectively, and some coupling between the mediators.
On the other hand, we are ignoring quartic couplings among the singlet and the mediators, and those where a Higgs boson is involved, since we are interested in the model with a $t/u$-channel mediator and not in the quartic quartic coupling channel -- or as a Higgs-portal model which have been extensively studied in the literature and has larger direct detection prospects. Moreover, for simplicity we take $Y_Q^{33} = 1$ and zero for other Baryonic couplings, so that our $Q_E$ field only decays to a top and bottom quark pair. This leaves us with five parameters, four of which are dimensionful. In principle $\lambda_2$ would also have an important impact on the annihilation of singlets to mediators, while changing the relationships of the quartic couplings may have some impact on the stability results. In future it would be interesting to perform a more sophisticated scan to allow for a more high-dimensional parameter space. 

To explore our parameter space, we performed a series of scans, starting from a uniform grid, then implementing several parallel Markov Chain Monte Carlo scans via the Metropolis-Hastings algorithm distributed across multiple cores on a cluster. Since we are interested only in the upper bound on the singlet mass, we construct a likelihood function $\mathcal{L}$ as a product:
\begin{align}
\mathcal{L} \equiv \mathcal{L}_{\rm upper}  (\Omega h^2, 0.120, 0.001 ) \times \mathcal{L}_{\rm upper} (a_0, 0.5, 0.001) \times \mathcal{L}_{\rm upper} (\delta_{\rm stability}, 1, 0.2)  \times  \mathcal{L}_{\rm bias} (m_S, \ov{m}_S, 0.2) 
\end{align}
where the first three likelihoods are sigmoid functions that cut off smoothly above the upper limit:
\begin{align}
\mathcal{L}_{\rm upper} (x, \ov{x}, s) \equiv \frac{1}{1 + \exp((x-\ov{x})/s)} 
\end{align}
and $\delta_{\rm stability}$ is $1$ for a stable vacuum and $0$ otherwise. This amounts to fixed large bias for stable over unstable points\footnote{In principle we could check for metastability and assign a likelihood based on a tunnelling probability. However, other than adding a significant complication, this is not very meaningful for this model since such points would correspond to large trilinear couplings and a loss of perturbativity.} without categorically excluding unstable points. The second term of the combined likelihood corresponds to the unitarity constraint. The last term is a bias on the dark matter mass, forcing the scan to probe heavier singlets:
\begin{align}
 \mathcal{L}_{\rm bias} (m_S, \ov{m}_S, s) = \left( \frac{m_S}{\ov{m}_S} \right)^{s}.
\end{align}
The value of $\ov{m}_S$ differs depending on the scan. After completing the MCMC scans, we select the points of the sample that strictly satisfy our constraints, which are therefore imposed as ``hard cuts''. Employing MCMC scans bears the advantage that a valid parameter space can be proposed more efficiently than a grid or random scan because the latter focus on regions that are allowed and avoid wasting computational resources on regions that are clearly excluded. In all MCMC scans, we select the largest partial wave amplitude to get a ``good" point.

\begin{figure}
\centering
\begin{subfigure}{.33\textwidth}
  \centering
  \includegraphics[width=\linewidth]{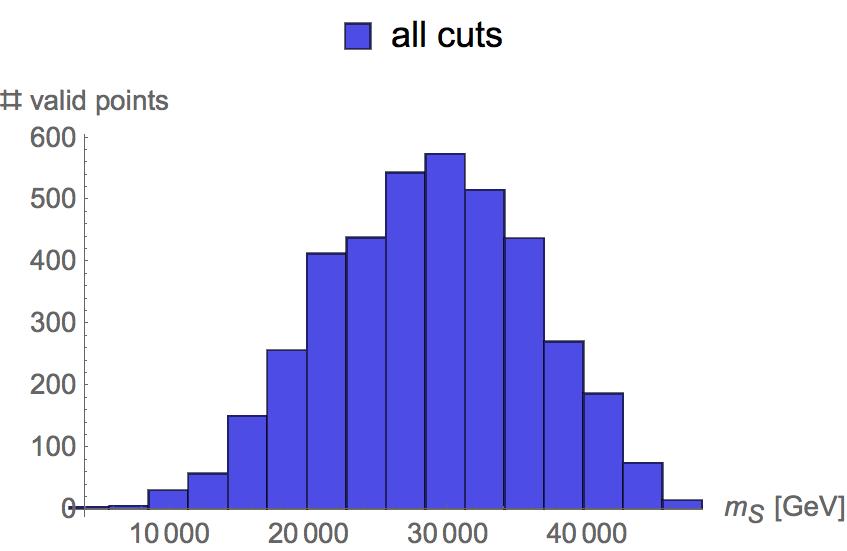}
  \caption{}
  \label{fig:onedAms}
\end{subfigure}%
\begin{subfigure}{.33\textwidth}
  \centering
  \includegraphics[width=\linewidth]{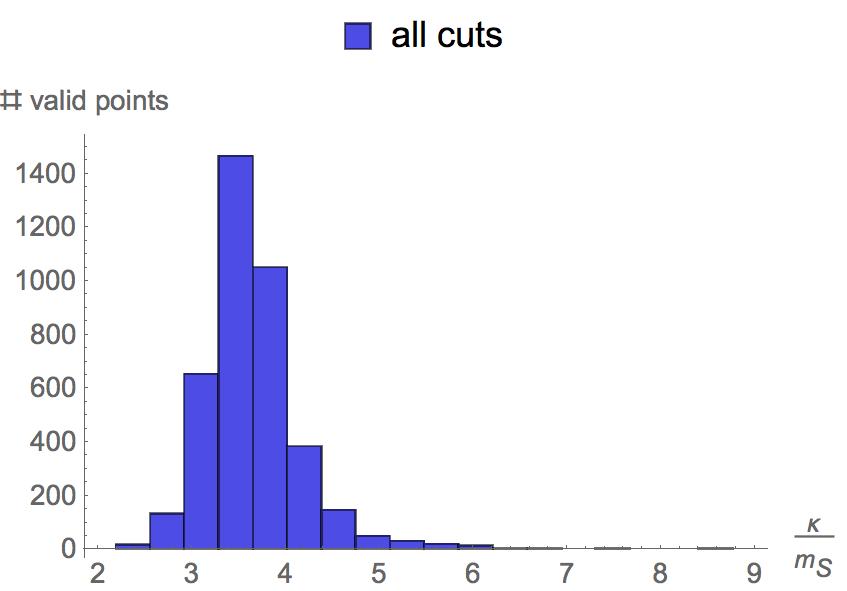}
  \caption{}
  \label{fig:onedAkappabyms}
\end{subfigure}
\begin{subfigure}{.33\textwidth}
  \centering
  \includegraphics[width=\linewidth]{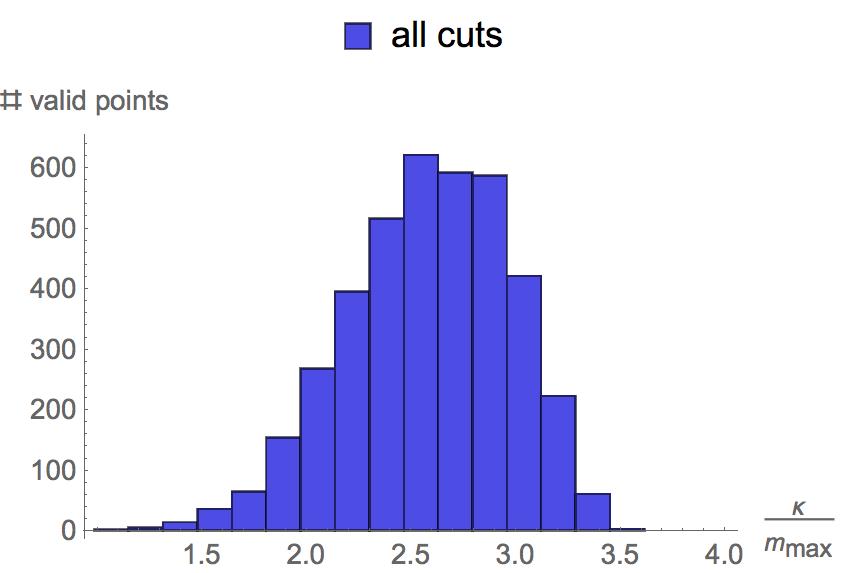}
  \caption{}
  \label{fig:onedAkappabym}
\end{subfigure}
\caption{Left: distribution of points as a function of $m_S$. There is a clear cutoff at $m_S\sim 47$ TeV. 
Middle: the ratio of the coupling $\kappa$ against $m_S$. While it peaks at around 3.5, there are values around 9, too.
Right: the ratio of $\kappa$ and the highest mass (being either $m_S$ or $m_{O}$). There is a clear cutoff at about 3.5, and a peak around 2.5.
The $y$-axis shows, in all three plots, how many of one million scan points made it through all cuts.
}
\label{fig:onedA}
\end{figure}

Figure~\ref{fig:onedAms} shows the distribution of the singlet mass after our scan, including only those points which passed all cuts.
In table~\ref{tab:Cutflow}, we list the amount of points that pass after each combination of cuts. 
Hereby, the cut on the mass hierarchy ensures that $m_S \leq m_O$, and that $\lambda_S \geq 0.5$.
There is a clear cutoff at around $m_S\simeq 47$ TeV, after which we found no more valid points. 
This implies a considerable amount of this mass range could be covered with a 100-TeV-collider.
This is also the central result of this paper.

\begin{table}[h]
\centering
\begin{tabular}{ll}
  \toprule
  Cut & number of points \\
  \midrule
  Mass hierarchy & 508918 \\
  Dark matter density (D) & 252098 \\
  Unitarity (U) & 359274 \\
  Vacuum stability (V) & 101365 \\
  U + D & 140163 \\
  U + V & 70056 \\
  D + V & 10568 \\
  All & 3963 \\
  \bottomrule
\end{tabular}
\caption{Points left over after each cut. The raw sample contained one million points. D refers to the cut on the dark matter density, U to that on unitarity, and V to that on vacuum stability. Details see text.}
\label{tab:Cutflow}
\end{table}

\begin{figure}
\centering
\begin{subfigure}{.33\textwidth}
  \centering
  \includegraphics[width=\linewidth]{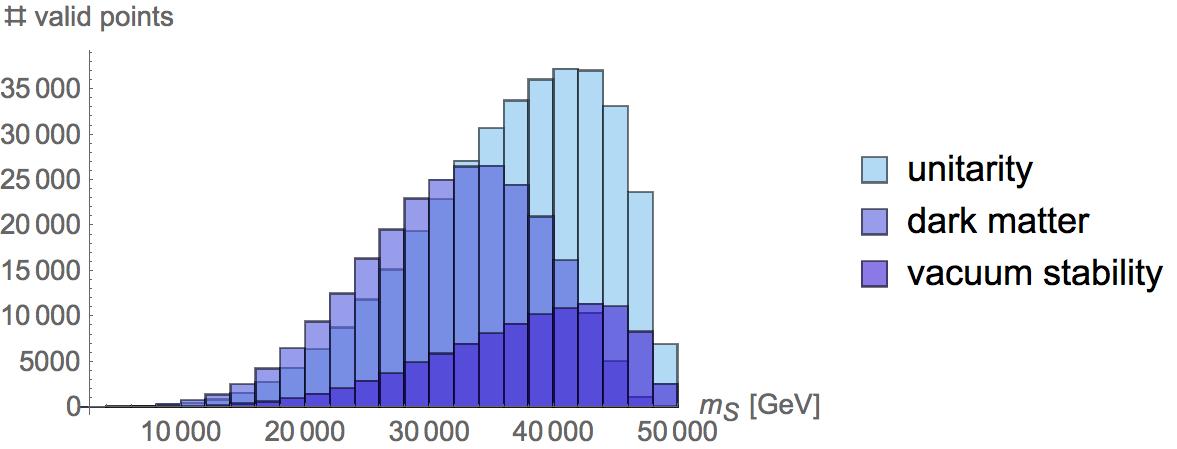}
  \caption{}
  \label{fig:onedUDVms}
\end{subfigure}%
\begin{subfigure}{.33\textwidth}
  \centering
  \includegraphics[width=\linewidth]{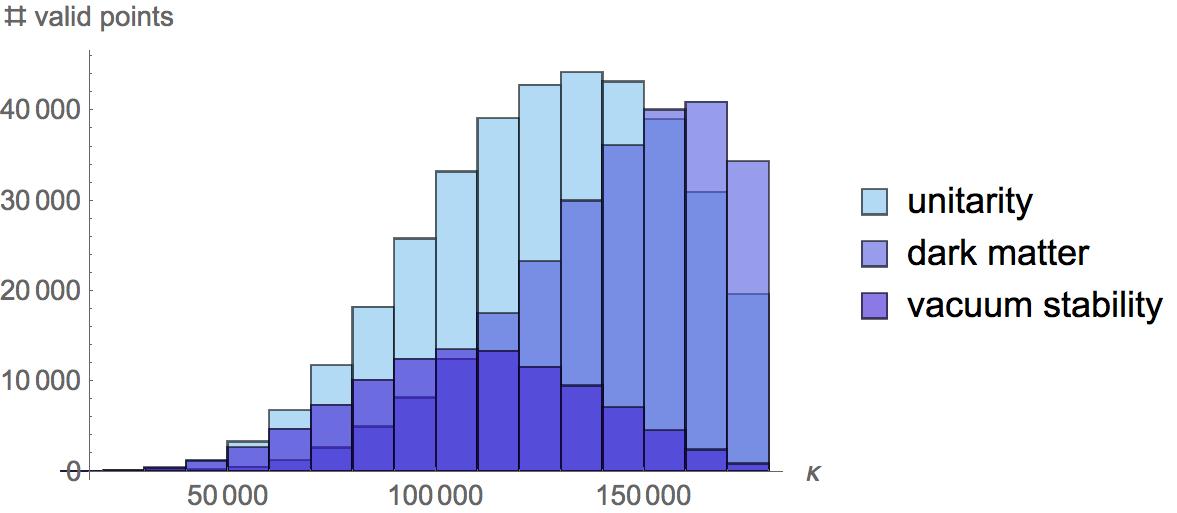}
  \caption{}
  \label{fig:onedUDVkappa}
\end{subfigure}
\begin{subfigure}{.33\textwidth}
  \centering
  \includegraphics[width=\linewidth]{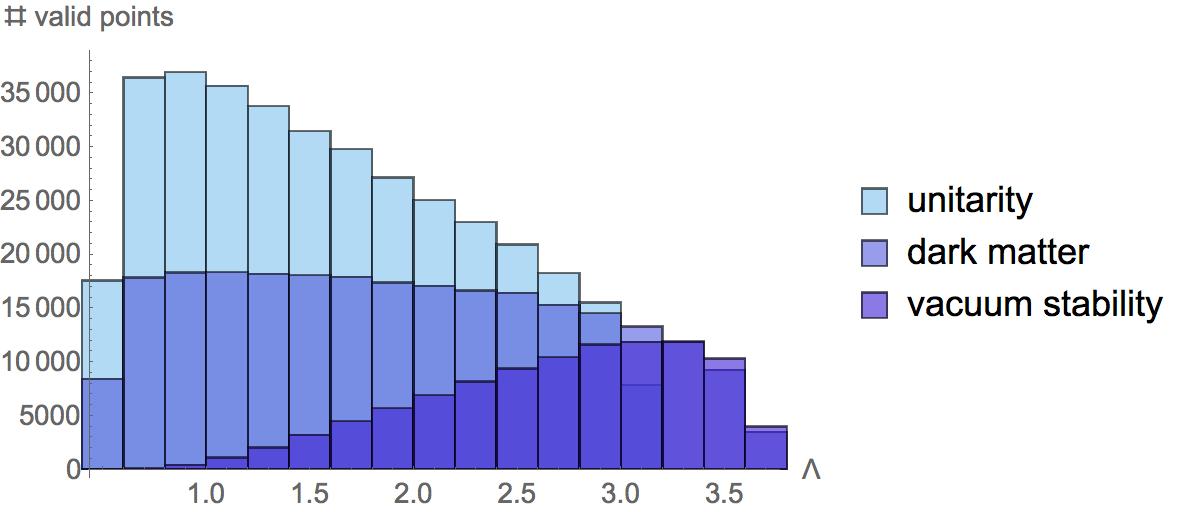}
  \caption{}
  \label{fig:onedUDVlams}
\end{subfigure}
\caption{Left: distribution of $m_S$ after various cuts.
Middle: the same for $\kappa$.
Right: the same for $\Lambda$. One can see that the cutoff at $\Lambda\sim \pi$ is due to unitarity.
The y-axis shows, in all three plots, how many of one million scan points made it through each cut, respectively.
In contrast to figure~\ref{fig:onedA}, these plots do not contain any information about which points make it through two or more cuts.
}
\label{fig:onedUDV}
\end{figure}

We show in figures~\ref{fig:onedUDVms}--\ref{fig:onedUDVlams} the effect of the separate cuts on the remaining points on the parameters $m_S, \kappa$ and $\Lambda$. We see that $\Lambda$ is bound by the naive unitarity constraint of $\pi$.

As was expected when setting up the model, we find a pretty clear relation between the strength of the coupling $\kappa$ and the masses of the involved particles.
This can be seen in figure~\ref{fig:onedAkappabyms}.
There is a clear peak around 3.5 for $\kappa/m_S$, although there are some outliers towards higher values.
If instead we take $\kappa$ in relation to the largest mass of each datapoint, i.e., one chooses the largest out of $m_S$ and $m_{O}$, the outliers disappear (figure~\ref{fig:onedAkappabym}). 
Instead, we find a peak at a ratio of about around 2.5.

\begin{figure}
\centering
\vspace{-1.0cm}
\begin{subfigure}{\textwidth}
  \centering
    \hspace{-6mm}
    \includegraphics[width=0.24\linewidth]{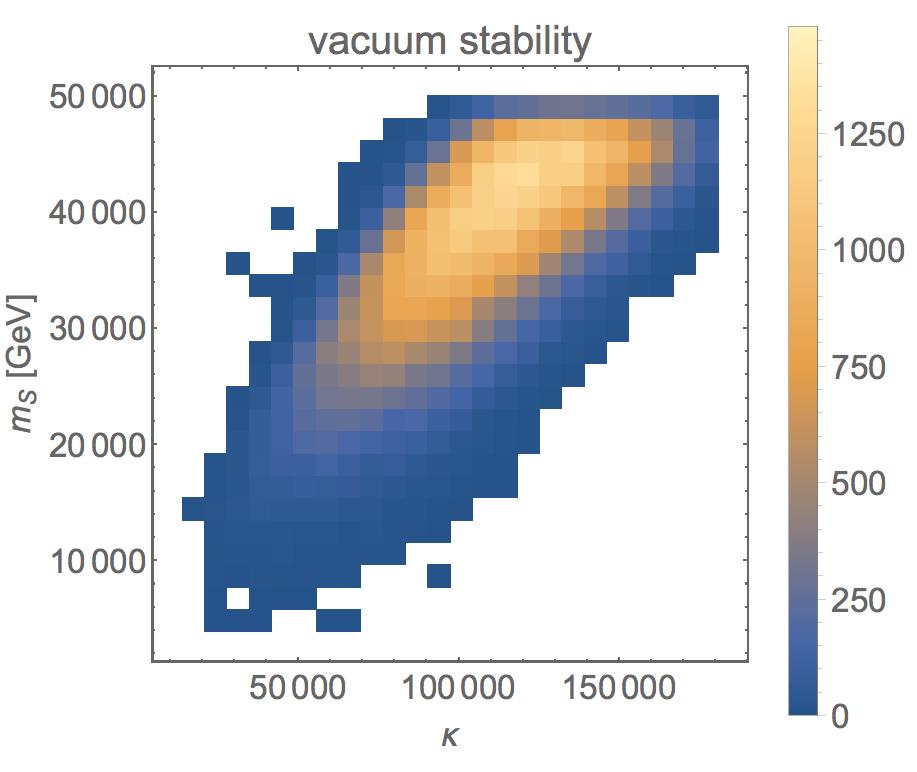}
    \includegraphics[width=0.24\linewidth]{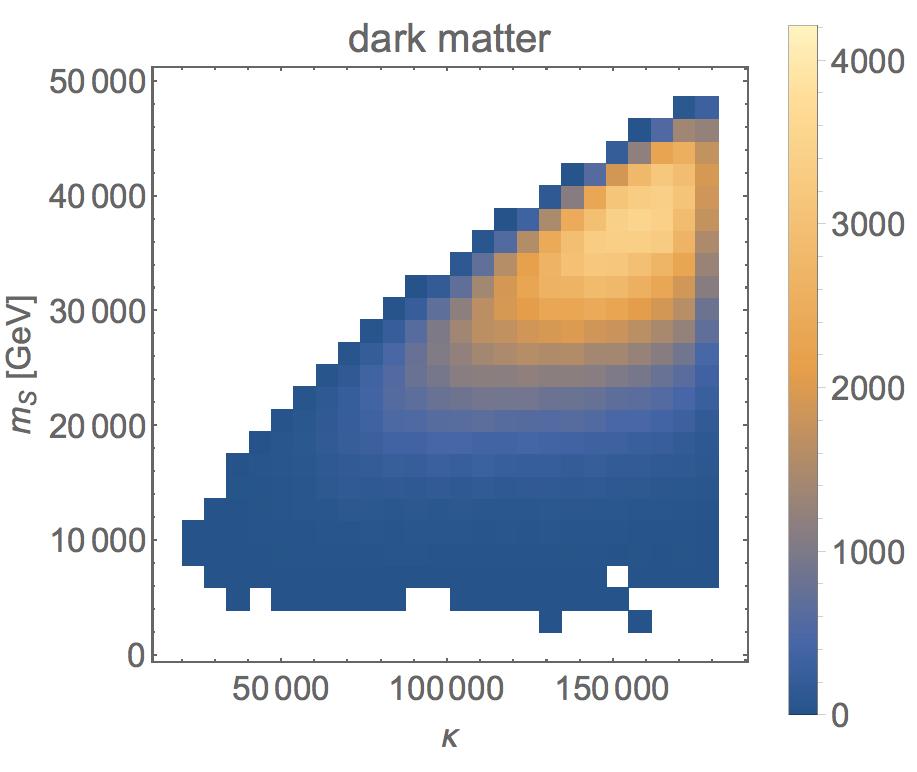}
    \includegraphics[width=0.24\linewidth]{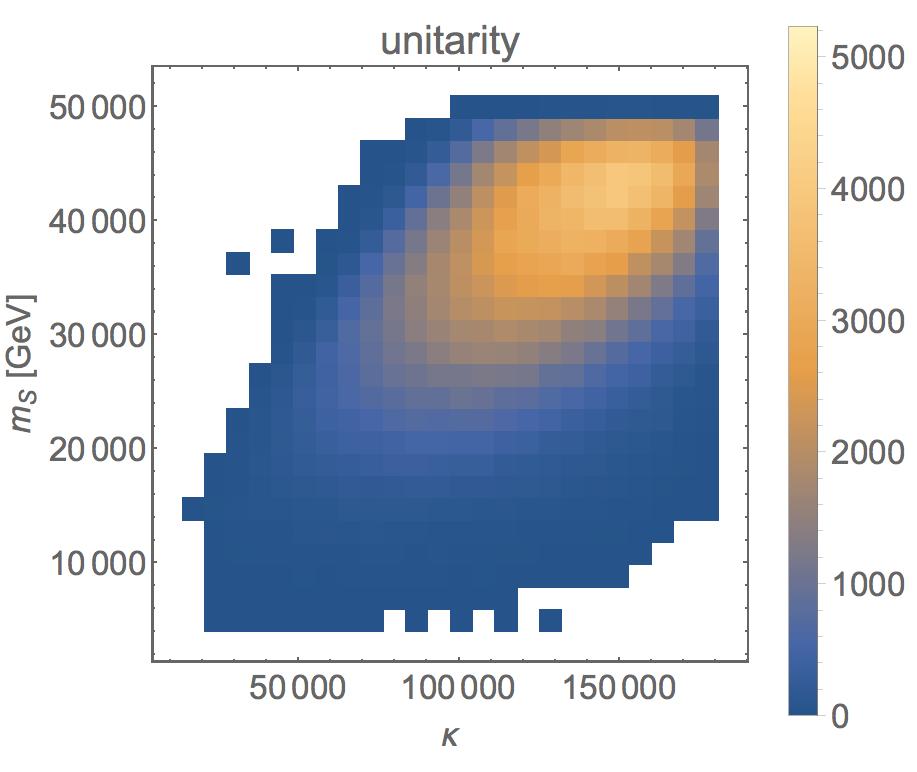}
    \includegraphics[width=0.24\linewidth]{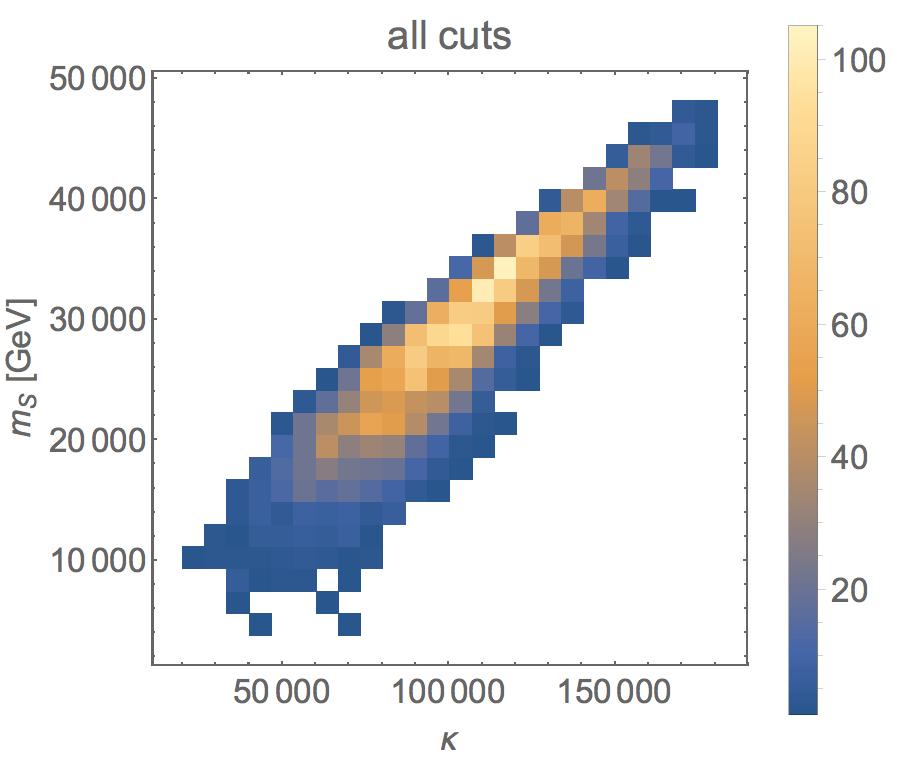}
  \caption{}
  \label{fig:twodUDVkappams}
\end{subfigure}
\begin{subfigure}{\textwidth}
  \centering
    \hspace{-6mm}
    \includegraphics[width=0.24\linewidth]{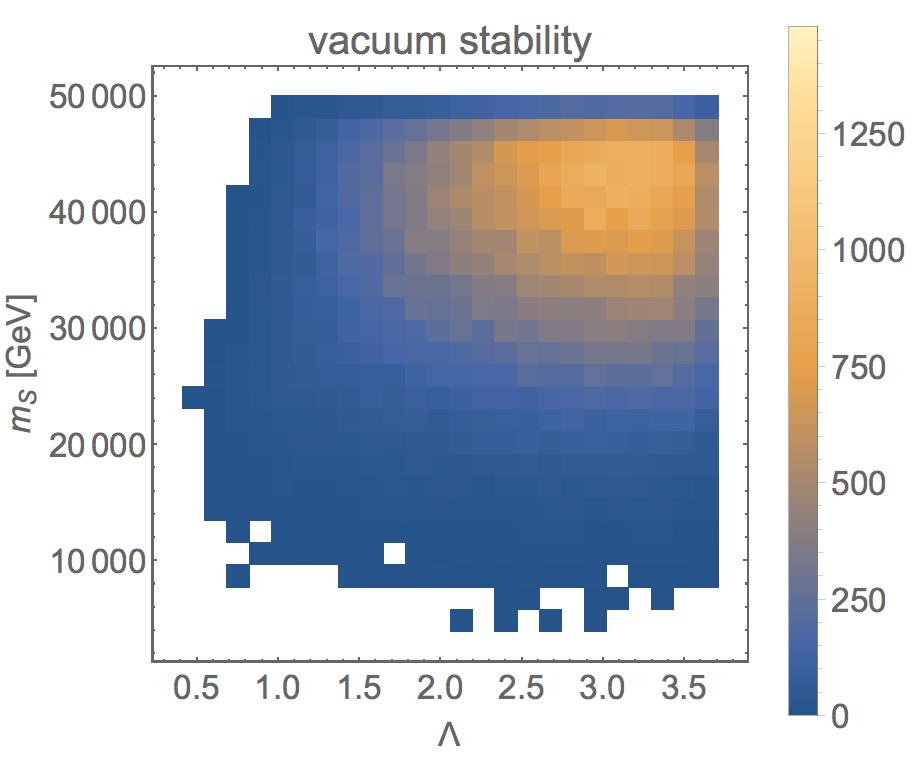}
    \includegraphics[width=0.24\linewidth]{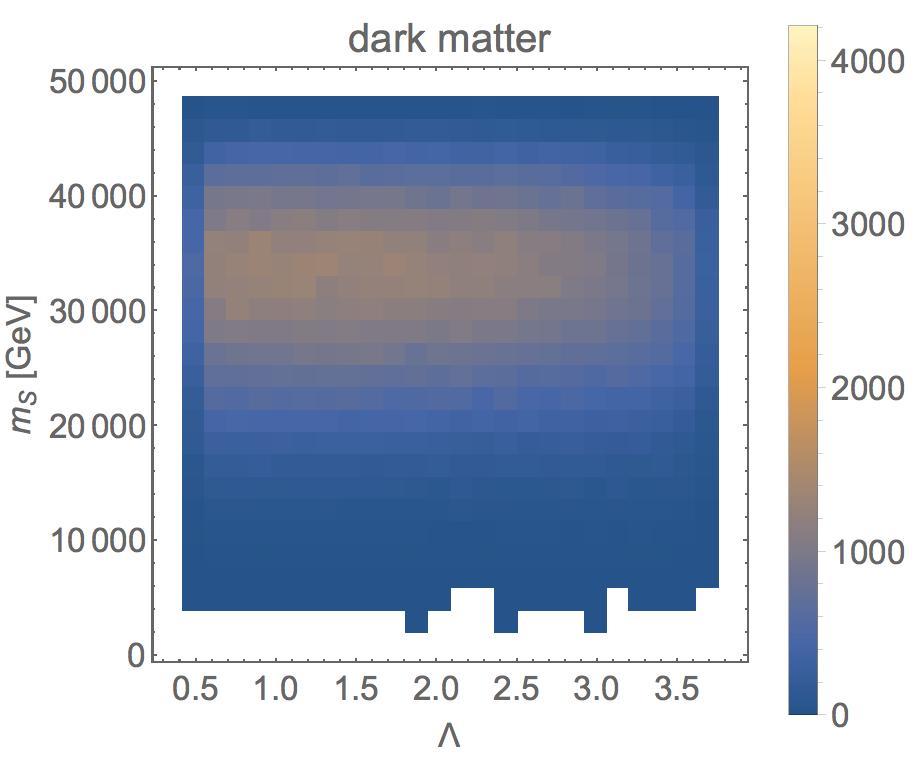}
    \includegraphics[width=0.24\linewidth]{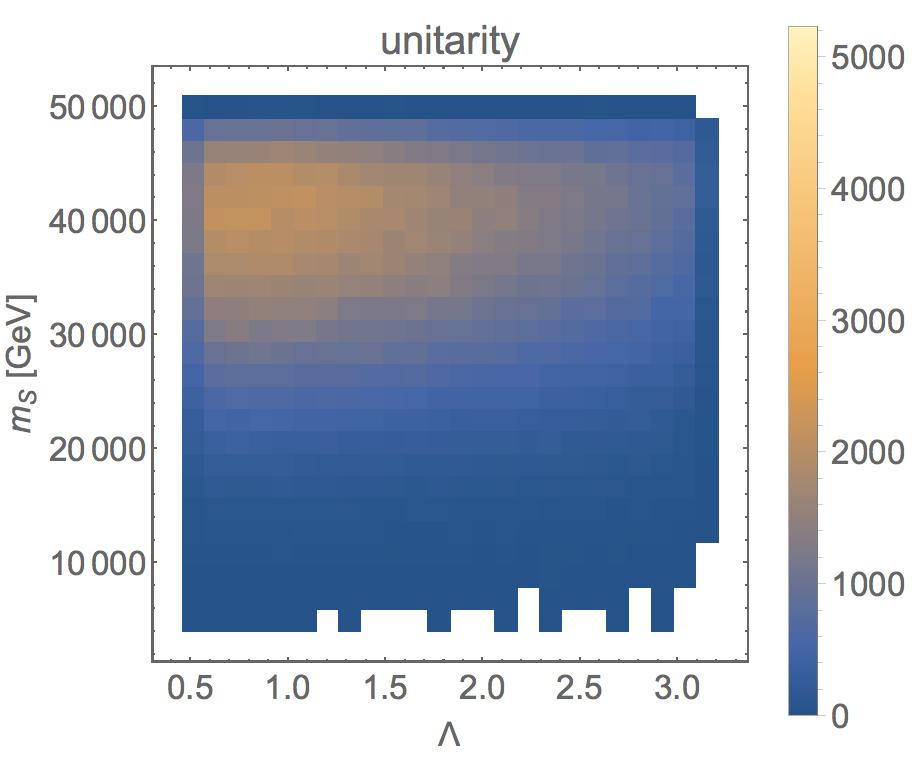}
    \includegraphics[width=0.24\linewidth]{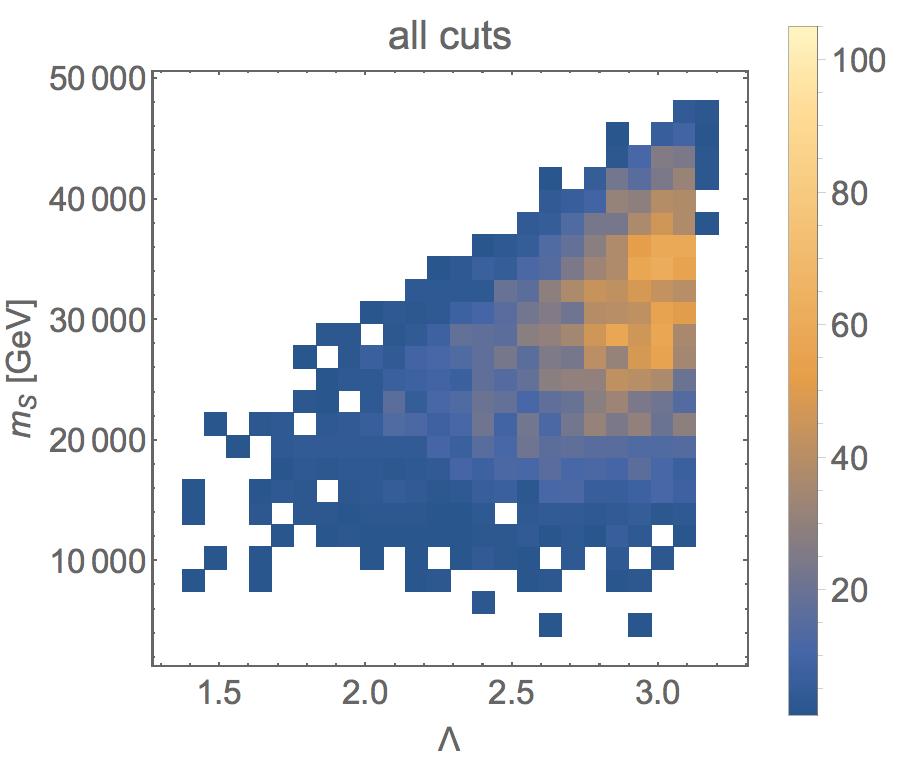}
  \caption{}
  \label{fig:twodUDVlamsms}
\end{subfigure}
\begin{subfigure}{\textwidth}
  \centering
    \hspace{-6mm}
    \includegraphics[width=0.24\linewidth]{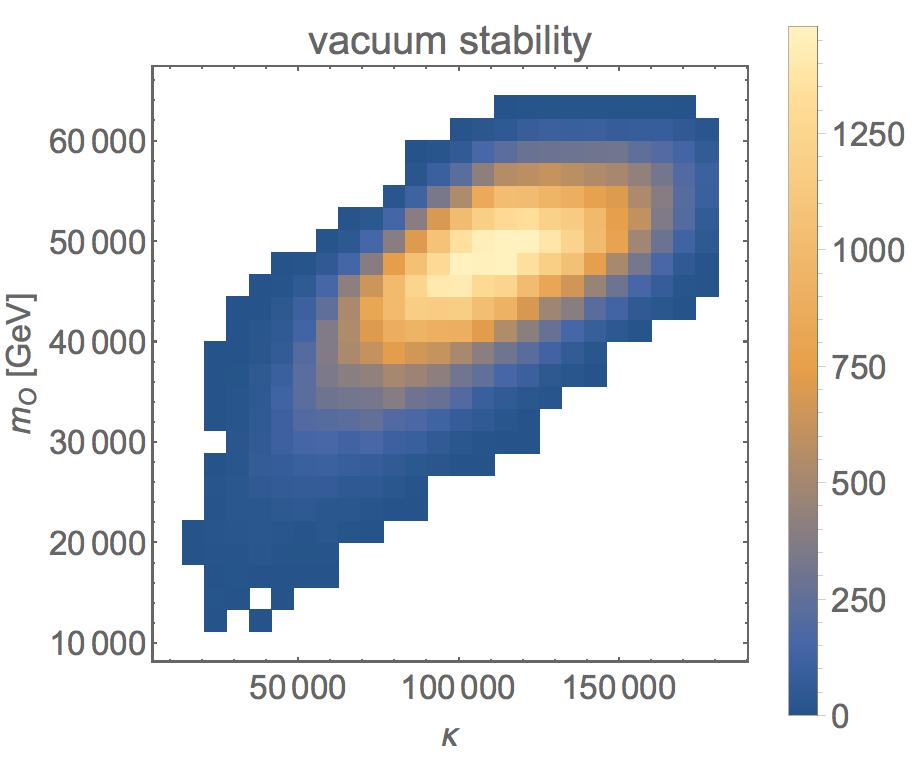}
    \includegraphics[width=0.24\linewidth]{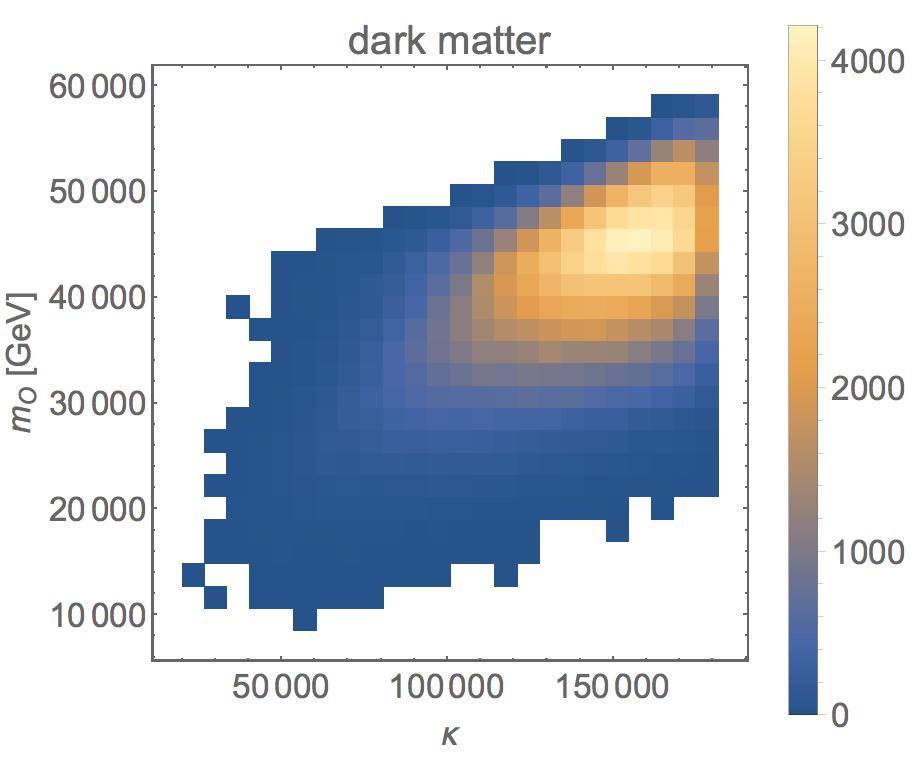}
    \includegraphics[width=0.24\linewidth]{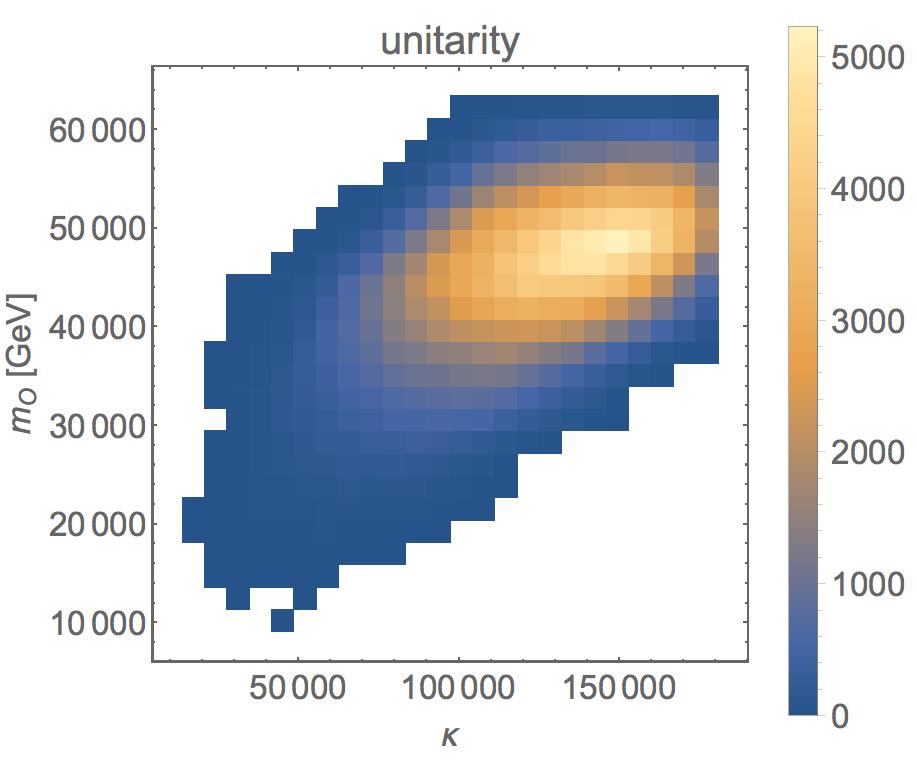}
    \includegraphics[width=0.24\linewidth]{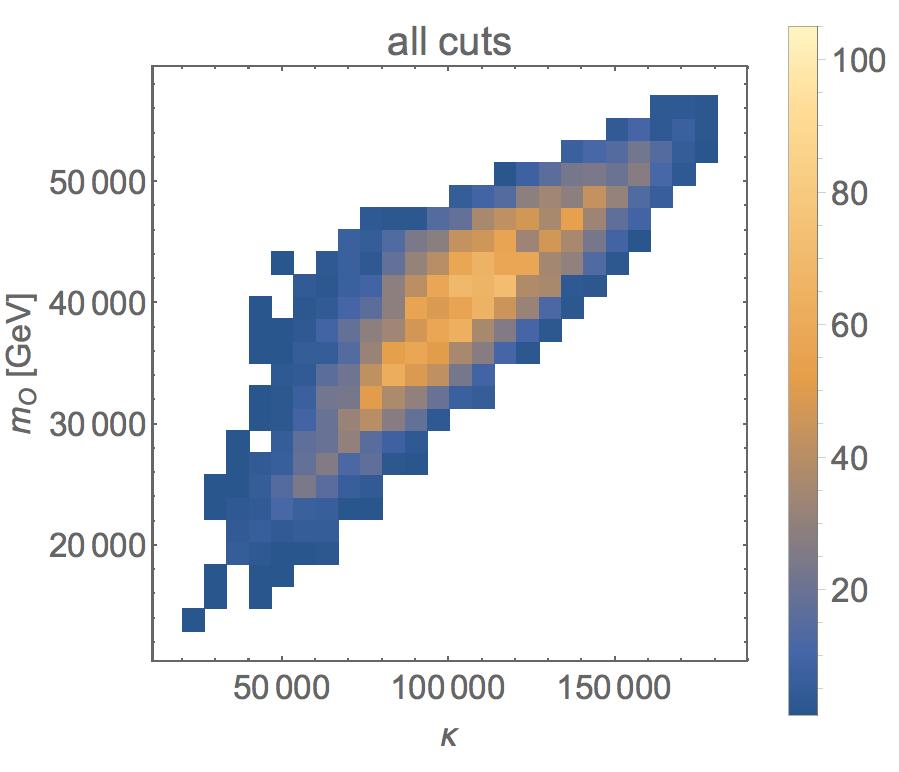}
  \caption{}
  \label{fig:twodUDVkappamqm}
\end{subfigure}
\caption{Left column: selected two-dimensional planes of the parameter space after cutting for vacuum stability.
Second  column: same, but cut for dark matter.
Third column: same, but cut for unitarity.
Right column: same, but after all three cuts.
Top: distribution of $\kappa$ against $m_S$. 
Middle: distribution of  $\Lambda$ against $m_S$. There is a clear cutoff at $\Lambda\simeq \pi$ due to unitarity.
Bottom: distribution of $\kappa$ against $m_O$, after each cut. As in~\ref{fig:onedUDVlams}, one sees that the cutoff at about $\pi$ is due to unitarity. 
The coloured regions indicate the regions where there were valid points after each cut, normalised for each plot (so a direct comparison of the colours between plots is not possible). 
}
\label{fig:twodUDV}
\end{figure}

Figure~\ref{fig:twodUDVkappams} shows the valid points in the $\kappa-m_S$-plane after each individual cut. One can see a clear correlation between the two, and the peak of $\kappa / m_S$ at 3.5  (figure~\ref{fig:onedAkappabyms}) is manifest. 
The outliers with a higher $\kappa / m_S$ ratio tend to be concentrated around the lower end of the distribution, where $\kappa$ is around 50 TeV and $m_S$ is below 10 TeV.
One can see that the vacuum stability constrains the allowed area from the bottom, i.e., the valid points are situated above the diagonal passing through $(\kappa, m_S) = (100 \mathrm{TeV},\ 10 \mathrm{TeV})$ and $(150 \mathrm{TeV},\ 30 \mathrm{TeV})$.
Likewise, the dark matter criterion constrains the allowed area from the top, i.e., the valid points are below the diagonal passing through $(50\mathrm{TeV},\ 25 \mathrm{TeV})$ and $(150 \mathrm{TeV},\ 45 \mathrm{TeV})$.

In figure~\ref{fig:twodUDVlamsms}, one can see the distribution of valid points in the $\Lambda-m_S$-plane after every cut.Vacuum stability eliminates points with low values of $\Lambda$ or $m_S$. The dark matter cut, by itself does not have much impact on the shape of the distribution. As expected, the cutoff $\Lambda\leq\pi$ is ensured by unitarity (third panel of figure~\ref{fig:twodUDVlamsms}). After all cuts, the points in the lower $m_S$ range are excluded, as expected, but also those above the diagonal passing through $(\Lambda, m_S) = (1.5,\ 25 \mathrm{TeV})$ and $(3.0,\ 50 \mathrm{TeV})$. The latter is a compound effect from the cuts on dark matter and unitarity, which shows that the dark matter cut does play a role after all.

Finally, figure~\ref{fig:twodUDVkappamqm} shows the distribution of valid points in the $\kappa-m_O$-plane. After each of the individual cuts, the resulting shape is bordered by three diagonals: one almost vertical one on the low-$\kappa$ end, and two more or less parallel ones going from the bottom left to the top right of the respective panel. The distribution of valid points after all cuts can be deduced almost directly from the overlap of the distributions after the three individual cuts.

\begin{figure}
\centering
  \begin{subfigure}{\textwidth}
    \centering
    \includegraphics[width=0.45\linewidth]{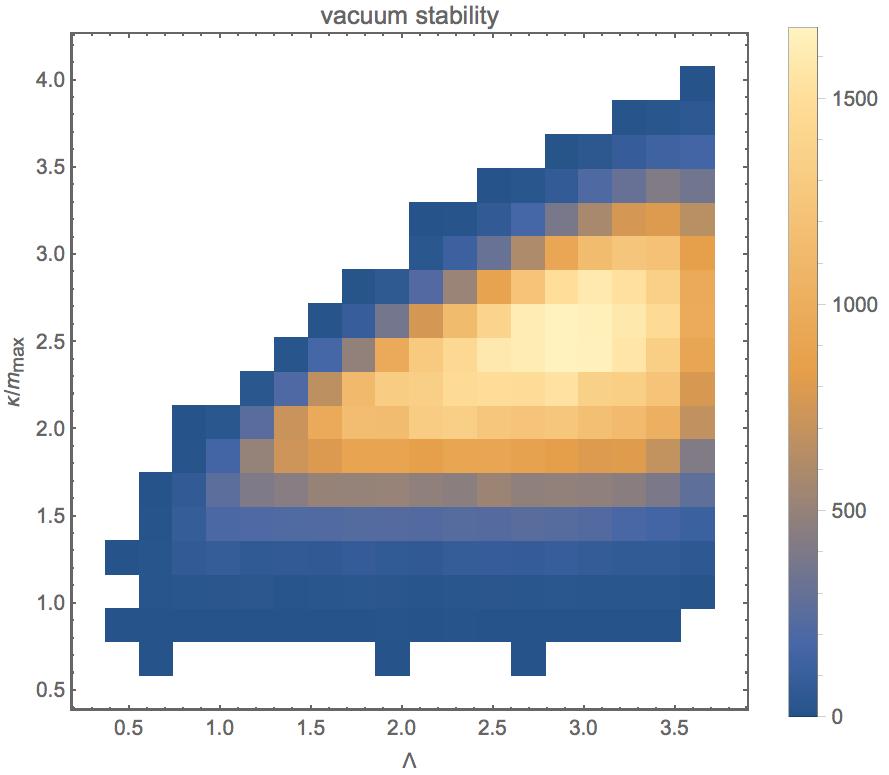}
    \hspace{5mm}
    \includegraphics[width=0.45\linewidth]{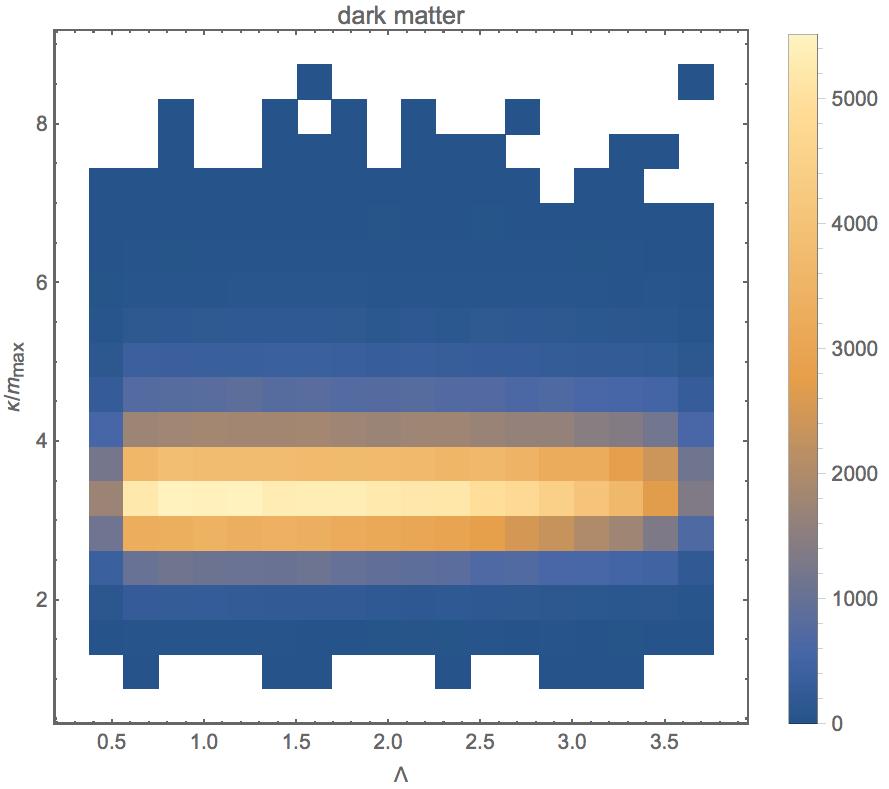}
  \end{subfigure}
  \begin{subfigure}{\textwidth}
    \centering
    \includegraphics[width=0.45\linewidth]{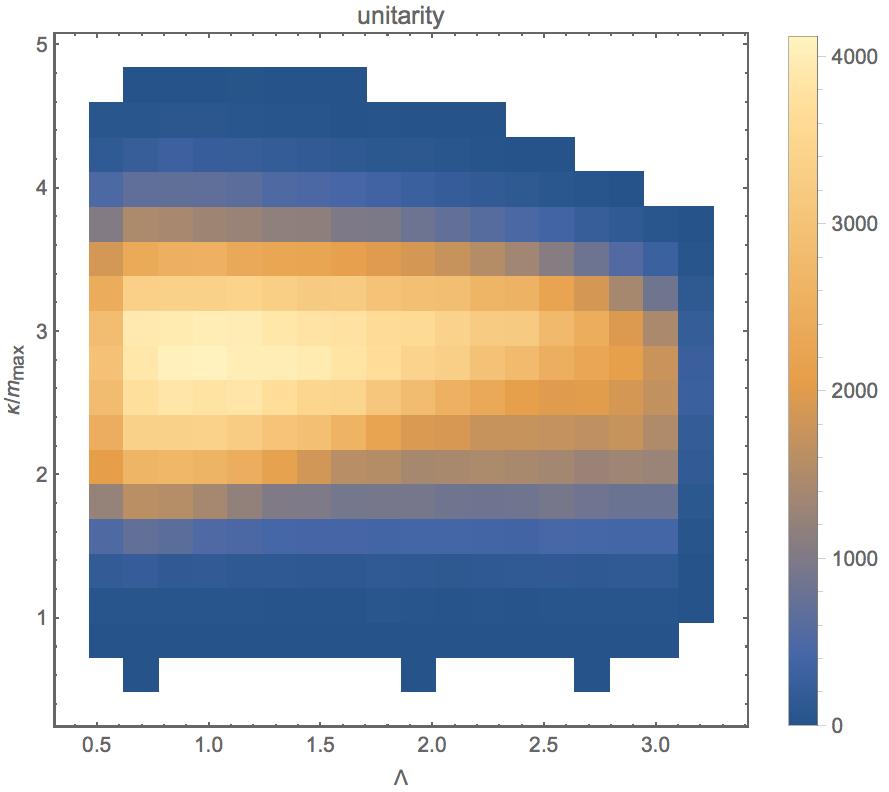}
    \hspace{5mm}
    \includegraphics[width=0.45\linewidth]{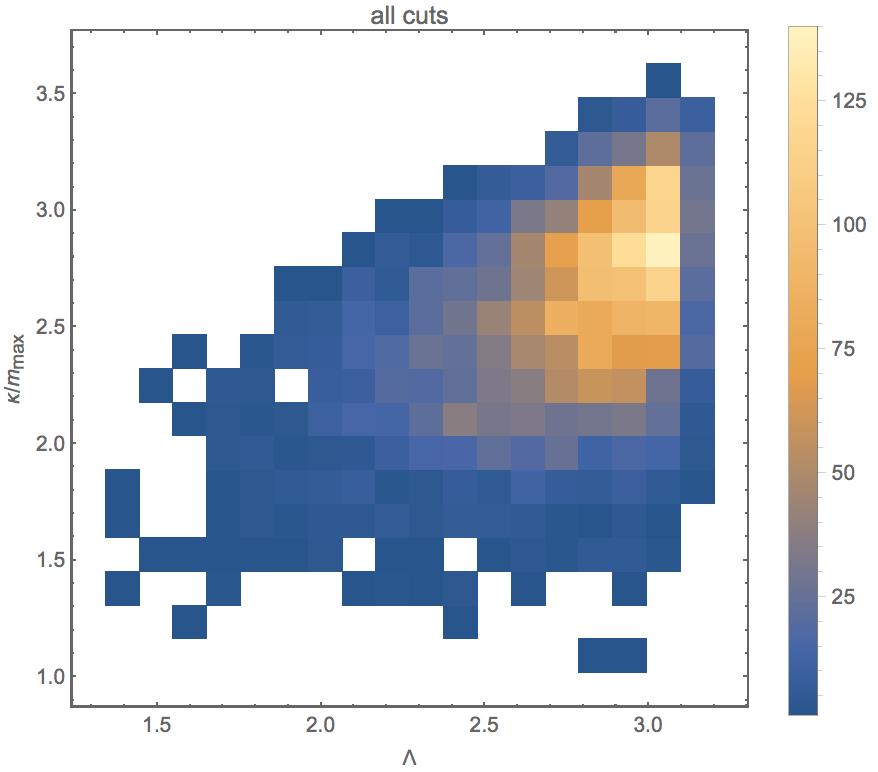}
  \end{subfigure}
\caption{Distribution of $\kappa$ over the biggest mass (from $m_S$ and $m_{O}$) against $\Lambda$.
Top left: valid points after the vacuum stability cut. Top right: same after dark matter cut. 
Bottom left: same after unitarity cuts. Bottom right: same after all cuts. 
For values above $\sim2$, one observes a linear relationship where $\kappa / m_\text{max}\simeq\Lambda$.
The z-axis shows how many of one million scan points made it through the cut(s).
The colour code is again normalised for each plot separately.
}
\label{fig:twodUDVAkappabymlams}
\end{figure}

Finally, figure~\ref{fig:twodUDVAkappabymlams} shows the distribution of $\kappa / m_\text{max}$ as a function of $\Lambda$ after various cuts.
One can see that vacuum stability imposes $\kappa  / m_\text{max} \lesssim \Lambda+1$.
Unitarity cuts away at some of the higher values of $\Lambda$ and $\kappa/ m_\text{max}$, and cuts off at $\Lambda \leq \pi$.

\subsection{Highest singlet mass}

The point that we found where the singlet was heaviest, and the main result of this section, has
\begin{align}
m_S = 47.354\, \TeV, m_O= 53.8747\, \TeV, m_E = 39.0254\, \TeV, \kappa = 174.121\, \TeV, \Lambda = 3.05993.
\end{align}
The dark matter relic density is $\Omega h^2 = 0.122$ for this point and the maximal $a_0 = 0.49$ is from the scattering matrix corresponding to the singlet representation (as might be expected from the earlier discussion), evaluated at $\sqrt{s} = 141\, \TeV,$ well away from any poles. This point is on the cusp of being ruled out by the unitarity calculation, which is dominated by the coupling $\kappa;$ we find that decreasing the coupling $\Lambda$ changes $a_0$ very little at this point but leads to an unstable vacuum already at $\Lambda=3,$ while increasing $\kappa$ to $180$ \TeV leads to $a_0 > 0.5$ (and also an unstable vacuum).

\subsection{Trilinears excluded by unitarity alone}

Finally we wish to highlight that, although most points conformed to the naive expectation that we could apply the limit $\Lambda < \pi$ from unitarity and constrain $\kappa$ just from vacuum stability, there are exceptions that underline the complementarity of the unitarity calculation. For example, 
\begin{align}
m_S =  26.07\,\TeV, m_O =  28.8\, \TeV,  m_E = 7.21\, \TeV,\kappa = 73.6\,\TeV, \Lambda =2.645 .
\end{align}
This point has $\Omega h^2 = 0.12$ and maximum $a_0 =0.51$ (again from the singlet submatrix) and the vacuum stability equations \emph{have no other solutions than the origin}. In fact, this point is typical of a whole branch of points where $ m_S \sim m_O \gg m_E$ for which this is true -- these points are excluded by unitarity because of the size of $\kappa$, but we would not have seen this either from classic unitarity bounds where $s \rightarrow \infty$ or from the vacuum stability constraints.

\section{Conclusions}
\label{SEC:CONCLUSIONS}

We have described the calculation and implementation of constraints from unitarity of scattering for $2 \rightarrow 2$ processes involving scalars of any representation under the strong gauge group, and finite scattering momentum. Since these unitarity constraints automatically constrain \emph{all} the scalar couplings of a theory, they are now very straightforward to include for a whole new class of models.

We also illustrated the utility of these routines and the complementarity of the information that they provide for studying dark matter models compared to vacuum stability and both naive infinite momentum perturbative unitarity constraints, and the ``absolute'' bound of ref.~\cite{Griest:1989wd}. We showed that there are points for which vacuum stability and ``naive'' unitarity are insufficient, i.e. the full perturbative unitarity calculation is indispensable. We introduced a toy model with a baryonic coupling and colourful mediators that decay in an interesting way to a top-bottom quark pair, that is a very simple example of the sort of models that can now be explored with these constraints. It would be very interesting to explore models with more complicated gauge representations.

The work also paves the way for several further extensions in future work: additional unbroken gauge groups; fermions and/or vectors in the scattering matrix; loop corrections. Moreover, our dark matter model had a maximum mass of $47\, \TeV,$ and coupled to colourful states, so much of the allowed parameter space would be accessible to a future $100\, \TeV$. It would therefore be interesting to consider dark matter-collider complementarity in terms of both its signatures at such a collider; but also the low-mass bounds at the LHC, since it could be searched for in the $t\ov{t} b \ov{b}$ channel.

\section*{Acknowledgments}

MDG acknowledges support from the grant
\mbox{``HiggsAutomator''} of the Agence Nationale de la Recherche
(ANR) (ANR-15-CE31-0002). He thanks Florian Staub for collaboration on including the colour factors in the unitarity routines in \SARAH. MDG also thanks Michael Baker for correspondence about those routines, and helping to identify bugs in the beta-version.

\appendix


\section{\SARAH implementation of our model: {\tt SM-SQQ}}
\label{APP:SARAHModel}

Since we implemented our model in \SARAH we list here the relevant parts of the model file (which is now also made public with version {\tt v4.14.4}). The new fields in addition to the SM are given as 
\begin{lstlisting}
ScalarFields[[2]] =  {s,  1, Sing, 0, 1, 1,-1};
ScalarFields[[3]] =  {qP, 1, QP, -1/3, 1, 3,1};
ScalarFields[[4]] =  {qM, 1, QM, -1/3, 1, 3,-1};
\end{lstlisting}
where the last line is the $\mathbb{Z}_2$ symmetry charge. So {\tt QP, QM} correspond to the fields $Q_E, Q_O$ in the body of the paper respectively. The Lagrangian is then given by the terms
\begin{lstlisting}
LagNoHC = -(mu2 conj[H].H + MS2/2 s.s  + MP2 qP.conj[qP] +MM2 qM.conj[qM] + LambdaS s.s.s.s + LambdaH conj[H].H.conj[H].H + LambdaHS/2  conj[H].H.s.s  + Lambda1/2 s.s.qM.conj[qM] + Lambda2/2 s.s.qP.conj[qP] + Lambda3 H.conj[H].qM.conj[qM] + Lambda4 H.conj[H].qP.conj[qP] + Lambda5 qP.conj[qP].qP.conj[qP] + Lambda6 qM.conj[qM].qM.conj[qM] + Lambda7 qP.conj[qP].qM.conj[qM]) 
LagHC = - ( Yd conj[H].d.q + Ye conj[H].e.l + Yu H.u.q + Kappa1 s.qP.conj[qM] + Yq qP.q.q + LambdaC/4 qP.conj[qM].qP.conj[qM] 
\end{lstlisting}
We did not explicitly give the colour structure for {\tt Lambda7} and hence we do not also include {\tt Lambda8} which has the same fields but a different contraction of the indices.

\subsection{Vacuum stability calculation}
\label{APP:VacStab}

Here we describe our routines for computing the vacuum stability constraints.

We begin with the potential for the fields $S, Q_E, Q_O$ where we define $S \equiv x, Q_E^1 \equiv \frac{1}{\sqrt{2}} y, Q_O^1 \equiv \frac{1}{\sqrt{2}} z$ for $x,y,z$ real and the other components of $Q_E, Q_O$ zero (since this is the most unstable direction in field space). This yields a potential
\begin{align}
  V =& \frac{1}{2} m_S^2 x^2 + \frac{1}{2} m_E^2 y^2 + \frac{1}{2} m_O^2 z^2 + \kappa xyz + \frac{1}{4} \lambda_1 x^2 y^2 + \frac{1}{4} \lambda_2 x^2 z^2 \nn\\
  & + \lambda_S x^4 + \frac{1}{4} \lambda_5 y^4 + \frac{1}{4} \lambda_6 z^4 + \frac{1}{4} \lambda_7 y^2 z^2 + \frac{1}{8} \mathrm{Re}(\lambda_C) y^2 z^2 
\end{align}
and then take the derivatives. These give three equations for which \emph{all} the solutions can be found with {\tt HOM4PS2}. However, we first rescale all of the dimensionful terms by
\begin{align}
m_S \rightarrow m_S/X, \quad m_E \rightarrow m_E/X,  m_O \rightarrow m_O/X, \kappa \rightarrow \kappa/X, \qquad X \equiv \mathrm{max} (\kappa, m_S, m_E, m_O).
\end{align}
We then calculate the numerical value of the potential at each of the solutions that we find and, if the minimum value is not at the origin of field space, we note that the vacuum is not stable.

\section{New routines in SARAH}
\label{SEC:SARAH}

With the release of version {\tt 4.14.4}, \SARAH contains updated routines to calculate unitarity constraints for scalars including colourful states (for now, no other unbroken non-abelian groups are considered for the unitarity routines). The algorithm used is as described in section \ref{SEC:ColourfulUnit}, with the group invariants hard-coded for certain common representations (in particular for octet representations the $f^{abc}$ and $d^{abc}$ matrices) but otherwise calculated by the included routines from {\tt Susyno} \cite{Fonseca:2011sy}. From the point of view of the user, the calculation functions exactly as in ref.~\cite{Goodsell:2018tti} except that colourful scalars are automatically included, unless they are explicitly removed from the scattering (as is done by default for many models). However, some new features have been added to aid performance/use/inspection of the results which will be described here.

\subsection{A new option for the cut-level of poles}

In \cite{Goodsell:2018tti}, several settings for the unitarity routines were outlined. For completeness, and to correct a misprint there, we give here the correct and updated complete options:

\begin{lstlisting}[language=SLHA,title=LesHouches.in.MODEL]
BLOCK SPhenoInput	 	 # 
 440 1               # Tree-level unitarity constraints (limit s->infinity) 
 441 1               # Full tree-level unitarity constraints 
 442 1000.           # sqrt(s_min)   
 443 2000.           # sqrt(s_max)   
 444 5               # steps   
 445 0               # running   
 446 2               # Cut-Level for poles
 447 0.25            # Tolerated relative proximity to s-channel poles
\end{lstlisting}
\begin{itemize}
 \item[{\tt 440}]: the tree-level unitarity constraints in the limit of large $\sqrt{s}$ can be turned on/off. Those include only the point interactions 
 \item[{\tt 441}]: the full tree-level calculations including propagator diagrams can be turned on/off.
 \item[{\tt 442}]: set the minimal scattering energy $\sqrt{s_{\rm min}}$ 
 \item[{\tt 443}]: set the maximal scattering energy $\sqrt{s_{\rm max}}$ 
 \item[{\tt 444}]: set the number of steps in which \SPheno should vary the scattering energy between $\sqrt{s_{\rm min}}$ and $\sqrt{s_{\rm max}}$. \SPheno will store the maximum eigenvalue. 
 For positive values, a linear distribution is used, for negative values a logarithmic one. 
 \item[{\tt 445}]: RGE running can be included to give an estimate of the higher order corrections
 \item[{\tt 446}]: How $t$ and $u$-channel poles are treated:
 \begin{itemize}
  \item[{\tt 0}]: no cut at all
  \item[{\tt 1}]: only the matrix element with a potential pole is dropped
  \item[{\tt 2}]: partial diagonalisation (default)
  \item[{\tt 3}]: entire irreducible sub-matrix is dropped
  \item[{\tt 4}]: {\bf disregard all unitarity constraints for this value of $s$}
 \end{itemize}
\item[{\tt 447}]: The relative proximity to $s$-channel poles that is allowed, $C_s$\footnote{See equation (11) of \cite{Goodsell:2018tti}.}.
\end{itemize}
With the new version, setting {\tt SPhenoInput 446} to $4$ will cause the unitarity constraints to be disregarded for the value of $s$ whenever a pole is found in an $s, t$ or $u$ channel of any diagram in \emph{any} scattering submatrix (the program will continue to scan over the range of values of $s$ in the hope of finding a valid constraint). This is the choice made in e.g. \cite{Betre:2014fva} and is the most conservative condition that can be placed, especially if coupled with a large $\sqrt{s_{\rm min}}$.

\subsection{Storage of symbolic form for each diagram}
\label{APP:Storage}

As part of the upgrade to the unitarity routines, couplings having more than one colour structure are properly taken into account and stored in new routines within {\tt SPhenoCouplings.f90}. While not all functionality of \SARAH will handle such couplings correctly yet (notably the loop decays) the unitarity routines and spectrum generation will give correct results. 

To enable cross-checks and reuse of the new routines, \SARAH writes a file {\tt Unitarity.m} which contains symbolic information about each scattering diagram computed. The format is
\begin{lstlisting}
{{s1,s2,s3,s4,prop,Type,colourrep,{dyn1,dyn2,dyn3,dyn4}},{couplings}}
\end{lstlisting}
{\tt prop} is the field appearing in the propagator (or just {\tt 1} for a quartic coupling); {\tt Type} is one of {\tt Q, S, T,U} meaning quartic, $s/t/u$-channel; {\tt dyn1, dyn2, dyn3, dyn4} are the dynkin indices of the fields {\tt s1,s2,s3,s4}. The fields are given as \emph{incoming} states, and the ordering is such that for $s$-channel and quartic interactions {\tt s1,s2 $\rightarrow$ s3,s4} while for the $t$-channel it is {\tt s1,s3 $\rightarrow$ s2,s4} and for $u$-channel it is {\tt s1,s3 $\rightarrow$ s4,s2}. Some typical lines for the model described in  this paper in the \SARAH notation of appendix \ref{APP:SARAHModel} would be
\begin{lstlisting}
{{QP, conj[QP], QM, conj[QM], hh, S, {0}, {{1, 0}, {0, 1}, {1, 0}, {0, 1}}},{{3*cp1[1]*cp2[1]}}},
{{QP, Sing, conj[QP], Sing, conj[QM], T, {0}, {{1, 0}, {0, 1}, {0}, {0}}},{{Sqrt[3]*cp1[1]*cp2[1]}}},
{{conj[QP], conj[QP], QP, QP, 1, Q, {1, 0}, {{0, 1}, {0, 1}, {1, 0}, {1, 0}}},{{-cp1[1] + cp1[2]}}},
{{conj[QP], conj[QP], QP, QP, 1, Q, {0, 2}, {{0, 1}, {0, 1}, {1, 0}, {1, 0}}},{{cp1[1] + cp1[2]}}},
\end{lstlisting}
where {\tt cp1[i], cp2[j]} are the first and second (if present) couplings in the diagram, and {\tt i,j} here refer to the number of the colour structure.  So for the first line, we have a colour-singlet scattering via an $s$-channel pole, where $Q_P, \ov{Q}_P \rightarrow Q_M, \ov{Q}_M$. The couplings in each case here are $- v \lambda_3$ and $- v \lambda_4$ and the overall colour factor is $\sqrt{3} \times \sqrt{3} = 3$, similar to the cases in section \ref{SEC:ColourExamples}. 
The seconnd line involves the scattering $Q_E , \ov{Q}_E \rightarrow S, S $ via the $t$-channel exhange of $\ov{Q}_M$ in the singlet representation. The colour factor here is again as derived in section \ref{SEC:ColourExamples}.

The third and fourth lines show a quartic coupling, so {\tt prop} is given as $1$ (since there is no propagator). However, the colour representations are $\mathbf{3}$ and $\mathbf{\ov{6}}$ respectively, and the two possible colour structures of the quartic coupling are involved. Recall that the lagrangian term is $-\lambda_5 |Q_E|^4$ so the two structures are $\delta_{ij} \delta_{kl}$ and $\delta_{il} \delta_{kj}$, both with coupling $-2 \lambda_5$ in this case. Hence we the net result for the antisymmetric term is $0$, and for the symmetric one it is $-4 \lambda_5$, exactly as we found in section \ref{SEC:ColourExamples}. 

One final note about the routines, to avoid confusion, is that, in \SARAH we actually calculate the matrix for $-a_0$ and then take the absolute values of the eigenvalues.

\subsection{Splitting into CP eigenstates}

With the inclusion of more fields, the program must find the eigenvalues of a larger scattering matrix and it is desirable to find simplifications where possible. In much the same way that we use the representations under charge and the strong force to decompose into scattering blocks, we can also use CP to reduce the rank of our matrices. If the user places the line
\begin{lstlisting}
UNITARITYCP=True;
\end{lstlisting}
in the file {\tt SPheno.m} for the model, \SARAH will attempt to assign CP charges for the states and decompose the scattering matrices accordingly. The user should find that the result is entirely unchanged, but for more complicated models some performance improvement may be found.

\bibliographystyle{h-physrev}
\bibliography{lit}

\end{document}